\documentclass[a4paper]{article}

\usepackage{graphicx}
\usepackage{latexsym}
\usepackage{amsmath}
\usepackage{amssymb}

\newenvironment{itemise}{\begin{itemize}}{\end{itemize}}
\newenvironment{centre}{\begin{center}}{\end{center}}

\newtheorem{proposition}{Proposition}

\newcommand{\labelbox}[1]{\parbox{10em}{\centering{#1}}}

\newcommand{\brak}[1]{\ensuremath{\left[#1\right]}}

\newcommand{\paren}[1]{\ensuremath{\left(#1\right)}}
\newcommand{\bparen}[1]{\ensuremath{\Bigl(#1\Bigr)}}

\newcommand{\abs}[1]{\ensuremath{\left|{#1}\right|}}
\newcommand{\floor}[1]{\ensuremath{\left\lfloor#1\right\rfloor}}

\newcommand{\goesto}{\rightarrow}
\newcommand{\half}{\frac{1}{2}}

\newcommand{\plog}[1][2]{\mathop{\mathrm{Li}_{#1}}}

\newcommand{\pref}{\succ}
\newcommand{\rema}{R_A}
\newcommand{\remb}{R_B}
\newcommand{\remc}{R_C}
\newcommand{\bshift}{\mathfrak{b}}
\newcommand{\rega}{\mathcal{K}_A}
\newcommand{\rarea}{\mathcal{A}}
\newcommand{\rarean}{\mathcal{A}_{\rema < 0}}
\newcommand{\rareap}{\mathcal{A}_{\rema > 0}}
\newcommand{\twarea}[1][]{\mathcal{A}_{\triwbl{#1}}}
\newcommand{\tlarea}[1][]{\mathcal{A}_{\trilbw{#1}}}
\newcommand{\tsarea}[1][]{\mathcal{A}_{\trisim{#1}}}
\newcommand{\triwbl}{\mathcal{T}}
\newcommand{\trilbw}{\mathcal{R}}
\newcommand{\trisim}{\mathcal{S}}
\newcommand{\typ}[2][1]{T_{#2|#1}}
\newcommand{\weight}[2]{\mathop{w}\paren{{#1},{#2}}}

\begin{document}

\title{The Australian Vote:\\Transferable Voting, Its Limitations and Strengths}
\author{Anthony B.\ Morton}
\date{July 2025}
\maketitle

\section{Introduction}
\label{sec:intro}

Australia, during the colonial era and in the decades after Federation in 1901, gained a reputation for democratic innovation.
This dates from at least the 1850s, when the then-novel practice of voting in secret in parliamentary elections was rapidly popularised and then exported to the rest of the world as `the Australian ballot'.
Though far less celebrated, the voting systems now known to theoreticians as Alternative Vote (AV) and Single Transferable Vote (STV) could justifiably be dubbed `the Australian vote'.
Even though these systems did not originate in Australia, they are in more widespread use across more levels of government in this country than in any other (in no small part due to the efforts of Federation-era advocates such as Catherine Helen Spence and Andrew Inglis Clark).
Australians generally refer to AV simply as `preferential voting', while the few other jurisdictions making use of this procedure (Ireland, Malta, individual US states) also refer to it as Ranked-Choice Voting or as Instant Runoff Voting.

Outside these jurisdictions, these two voting systems (strictly speaking, family of systems in the case of STV) are most commonly encountered in occasional political debates as the primary contenders to replace simple Plurality voting, known more popularly as first-past-the-post (FPTP).
The deficiencies of FPTP are well-known, and this paper both recaps the arguments and offers some new perspectives as to why AV (for single-member electorates) and STV (for multi-member electorates) does a much better job of expressing basic principles of `majority rule' and `voter choice' than other popular voting systems including FPTP.

All the same, AV and STV are vulnerable, like all `fair' voting systems, to paradoxical outcomes---scenarios where the procedures sometimes yield results that fail on compelling criteria of `fairness'.
Social-choice theorists, who study voting systems in detail, have tended to take a dim view of both AV and STV because these systems fail on a greater number of hypothetical criteria than certain alternative systems that theorists are inclined to favour \cite{f:rpapfesc,n:ortrvsc}.
The fact that none of these alternative systems are in use in real elections, even while FPTP, AV and other `inferior' systems persist, may be cause to reconsider \cite{rg:pppidav}.
It is a purpose of this paper to provide considerations, some of them novel, by which a reasonable person might be led to favour an `inferior' system like AV or STV for elections at-large even among populations well acquainted with social-choice theory.

The considerations in this paper are informed by the author's own extensive experience of Australian elections and study of the outcomes in practice of AV and STV (here referred to collectively as \emph{transferable voting}).
It is thus an \emph{apologia} of sorts, but offered in the hope it helps develop and reveal the internal logic of these transferable voting systems in ways that have perhaps benefited from this experience.

\section{Fundamentals and Nomenclature}
\label{sec:nom}

\subsection{Transferable voting}
\label{sec:voting}

AV and STV are essentially variants of the same system, originally formulated (in its STV variant that elects multiple winners) independently by the British lawyer Thomas Hare in 1857 and by the Danish politician and mathematician Carl George Andr\ae\ in 1855 (although the English schoolmaster Thomas Wright Hill in 1819 should also be noted as a precursor).
The essence of this \emph{transferable voting} system, whether used to select one winner or a prescribed number $W$ of winners, is as follows:
\begin{enumerate}
\item
Each individual voter ranks the candidates in their chosen order of preference.
(In some variants the ranking must extend to all or practically all candidates, while in others a voter may choose not to rank those beyond an initial preferred set.)
\item
A \emph{quota} $Q$ of votes is calculated such that no more than $W$ candidates can receive more than $Q$ votes each.
Any candidate that attains $Q$ votes at any stage of counting is elected.
\item
If no candidate attains $Q$ votes, candidates are eliminated one by one until some non-eliminated candidate attains $Q$ votes.
\item
The candidate eliminated at each stage is the one with the lowest number of first-place votes out of those candidates not yet elected or eliminated.
(In the rare case of a tie for the lowest number, candidates may be eliminated simultaneously---but see below.)
\item
Each vote for an eliminated candidate is `transferred' to the voter's next preferred candidate, in the sense that the voter at each stage is regarded as having voted for the candidate placed first out of those not yet elected or eliminated.
\item
In multi-winner elections, elected candidates are eliminated from the count at subsequent stages and their `surplus' votes in excess of the quota $Q$ are transferred as for eliminated candidates (with various different methods used to account for this---see below).
Surplus transfers occur before any further candidates are eliminated, to ensure voter preferences are recognised.
\item
Where a voter does not rank all candidates, but all of their ranked candidates are either elected or eliminated, the vote is considered `exhausted' and no longer counts.
The quota $Q$ may or may not be recalculated at each stage to account for the reduced total due to exhausted votes.
\end{enumerate}
Another way to describe the operation of the system is as a multi-stage process, at each stage of which candidates are classified as either \emph{elected}, \emph{eliminated}, or (the remainder) \emph{contesting}.
All candidates start out as contesting, with candidates who are elected or eliminated remaining so permanently, and the number of contesting candidates dropping with each stage (with at least one either elected or eliminated each time) until there are $W$ elected candidates.

The ranking of candidates returned by each voter is also known as a \emph{ballot}.
The term `vote' is sometimes taken to refer just to the entry against a particular candidate on a ballot, but here the terms `ballot' and `vote' will be used interchangeably without a formal distinction being made.

\subsection{Quota}
\label{sec:quota}

Specific STV systems differ not only in the number of winners $W$, but also in the manner in which the quota $Q$ is calculated and the way in which surplus votes are transferred.
The most common method of calculating $Q$ is as the so-called \emph{Droop quota}, where if $N$ is the total number of formal votes, then
\begin{equation}
Q = \floor{\frac{N}{W + 1}} + 1,
\label{eq:droop}
\end{equation}
where $\floor{x}$ is the floor function, the greatest integer not exceeding $x$.
The quota $Q$ by this formula is the \emph{smallest} whole number of votes that can be received by each of $W$ winning candidates but not by $W + 1$ candidates.
(The original \emph{Hare quota} $\floor{N/W}$ by contrast is the \emph{largest} whole number of votes able to be received by each of $W$ candidates; it is nowadays less favoured for STV elections because it complicates the count at later stages without necessarily providing greater proportionality.)

When considering closely contested elections it can be useful to have an exact formula for $Q$, which can be obtained from the exact division formula for $N / (W + 1)$, namely
\begin{equation}
N = Q' (W + 1) + r, \qquad 0 \leq r \leq W,
\label{eq:ndivw}
\end{equation}
which uniquely defines the quotient $Q'$ (with $Q = Q' + 1$) and the remainder $r$ as an integer between 0 and $W$.
This leads to the exact formula corresponding to (\ref{eq:droop}):
\begin{equation}
Q = \frac{N + \rho}{W + 1}, \qquad \rho \in [1, W + 1]
\label{eq:quota}
\end{equation}
where $\rho = W + 1 - r$ is the unique integer between 1 and $W + 1$ that when added to $N$ makes the result exactly divisible by $W + 1$.
(So if $N$ itself is divisible by $W + 1$ then $\rho = W + 1$, not zero.)

If voters do not rank all candidates, then votes may exhaust during the count and the number $N$ will shrink at each stage by the number of exhausted votes, which in turn will cause $Q$ to shrink at least notionally.
($Q$ may be recalculated in actuality but usually is not---see below.)

In a single-winner election, $W = 1$, the process is complete once a single candidate attains the quota $Q = \floor{N/2} + 1$, which is termed the \emph{absolute majority} from $N$ ballots.
As there can only be one winner there is no need to transfer surplus votes to determine other winners, and the process reduces to one of successive elimination until one candidate attains an absolute majority of original and transferred votes.
This is the system known as Alternative Vote (AV).

\subsection{Surplus transfer votes and Hare--Clark STV}
\label{sec:hareclark}

In multi-winner elections, at the point where a candidate is elected and further winning places remain to be filled, the `excess' of actual votes above the candidate's quota $Q$ must be considered; otherwise, there would be insufficient remaining ballots to determine additional winners on an equal-value basis.
If an elected candidate $X$ has received $x \geq Q$ votes, a fair process should remove the equivalent of $Q$ ballots from the count, and reallocate the \emph{surplus} $s = x - Q$ to other contesting candidates.
(The exception is the fortunate but rare case $x = Q$, when nothing further has to be done, and the process continues as though electing $W - 1$ winners from the remaining $N - Q$ ballots---note that the same quota $Q$ may fairly be used for this next stage, as will be stated more precisely below.)

In Hare and Andr\ae's original STV system, the surplus $s$ is transferred by sampling $s$ ballots at random from the $x$ ballots cast for $X$, and transferring each of these to their next preferred candidate.
This system is still in use in local elections in Cambridge, Massachusetts, but others consider it unsatisfactory, as it introduces an element of chance into elections.
This problem led to the emergence of the alternative \emph{Gregory method}, as first applied to elections in Tasmania by then Attorney-General Andrew Inglis Clark in 1896.
This system involves associating a \emph{value} to each ballot, with all ballots initially having value 1, then when candidate $X$ is elected reducing all $x$ votes for $X$ in value by the fraction
\begin{equation}
v_x = \frac{s}{x} = 1 - \frac{Q}{x}.
\label{eq:gregory}
\end{equation}
This reduction in value ensures that \emph{all} votes for $X$ can be transferred, but scaled so as to effectively remove $Q$ votes' worth of value from the count.
The process is applied recursively, albeit with some technical variations, as further candidates are elected.
This form of STV, that uses the Gregory fractional method to transfer the surplus for elected candidates, is known as the \emph{Hare--Clark} system.

Hare--Clark STV is used for parliamentary elections in Tasmania and the Australian Capital Territory.
The STV system used for Australian Senate elections is based on Hare--Clark but supplements it with an `above the line' option that mimics a party-list system.
Prior to 2016, voters for the Senate could mark a single vote `above the line' which would direct all their ranked preferences according to a closed party list (but one which orders all candidates).
This was replaced in 2016 with a system where voters themselves could rank parties `above the line' and this acts as a party-by-party shorthand for a ranked list of each party's `below the line' candidates.
Once the ranked party lists are notionally substituted, the system becomes functionally identical to Hare--Clark `below the line'.

All these Hare--Clark derived systems carry the possibility of exhausted ballots as voters are not required to rank all parties or candidates.
Exhausted ballots when transferring votes from elected candidates are dealt with in the Hare--Clark system by replacing the scaling $s / x$ according to (\ref{eq:gregory}) with $s / (x - z)$, where $z$ is the number (or strictly speaking, value) of exhausted ballots.
If $x - z \leq s$, equivalently $z \geq Q$, then no scaling is done.

Because systems such as Hare--Clark generally result in fractional vote values for candidates after the first is elected, it has been argued that the rounding of $Q$ to an integer as in (\ref{eq:droop}) is no longer justified, particularly as computers can now quickly execute the entire count in a verifiable manner.
Rounding has however been maintained in the definition of the Droop quota for Australian elections.
The choice to maintain the definition of $Q$ as an integer can be defended on the basis that it permits this decisive quantity to be stated precisely in electoral returns, and that when determining the election of a candidate there is no reliance on comparisons of two non-integer fractional quantities which might be arbitrarily close to one another.
Maintaining $Q$ as an integer can mean that in improbably close situations, a candidate fails to be elected who would be if a fractional quota were to be permitted; but as $Q$ itself is determined from a `minimality' criterion, this need not be considered an unfair outcome.

\subsection{Quota stability in multi-winner STV}
\label{sec:stability}

The rounding implicit in computation of the quota $Q$ via (\ref{eq:droop}) does introduce complications, even in the absence of exhausted ballots, because it means that $Q$, if notionally or actually recalculated at each stage of the count, will fluctuate slightly as the number (or in Hare--Clark, value) of ballots is reduced with each elected candidate in an STV election.

This fluctation in $Q$ is provably insignificant, as will be seen presently, and means that in practice recalculating $Q$ after electing candidates is not essential.
But it is of some interest to see how this works out mathematically, as this stability in $Q$ with number of elected candidates helps underlie STV's claims of \emph{proportional representation} in its outcomes.

Suppose that at some stage of the count the total number of ballots is $N_0$ (or in the case of Hare--Clark, the total equivalent value of ballots is $N_0$), and there are $W_0 > 1$ winners yet to be elected.
The quota calculated by (\ref{eq:droop}) from $N_0$ and $W_0$ is denoted $Q_0$.
Upon counting the votes, it is found that $e \in [1, W_0 - 1]$ candidates have each achieved a quota $Q_0$ of votes and are therefore elected (with or without a surplus to transfer).
Since $e < W_0$, there will be at least one further stage of counting with $W_e = W_0 - e$ winners left to elect.
It is assumed for now that no ballots exhaust.

By the exact quota formula (\ref{eq:quota}), one has
\begin{equation}
Q_0 = \frac{N_0 + \rho_0}{W_0 + 1}, \qquad \rho_0 \in [1, W_0 +1],
\label{eq:quota0}
\end{equation}
where again $\rho_0$ is uniquely determined from a division remainder.
After electing $e$ candidates, and assuming no ballots exhaust, the new reduced number (or value) of ballots is
\begin{equation}
N_e = N_0 - e Q_0,
\label{eq:nume0}
\end{equation}
or after substituting (\ref{eq:quota0}) for $Q_0$,
\begin{equation}
N_e = \frac{(W_0 + 1 - e) N_0 - e \rho_0}{W_0 + 1}.
\label{eq:nume}
\end{equation}
One may confirm that the numerator of (\ref{eq:nume}) is divisible by $W_0 + 1$: observe that $N_0 + \rho_0$ is so divisible by definition of $\rho_0$, and that multiplying $N_0 + \rho_0$ by the integer $(W_0 + 1 - e)$ yields a quantity that differs from the numerator of (\ref{eq:nume}) by a multiple of $W_0 + 1$.

Now multiply (\ref{eq:nume}) through by $W_0 + 1$ to clear the fraction, and subtract the quantity $(W_0 + 1 - e) N_e$ from both sides.
This results in
\begin{equation}
e N_e = (W_0 + 1 - e) (N_0 - N_e) - e \rho_0.
\label{eq:numee}
\end{equation}
Using (\ref{eq:nume0}) once again to write $(N_0 - N_e) = e Q_0$, and dividing through by $e \neq 0$, one finally obtains
\begin{equation}
N_e = (W_0 + 1 - e) Q_0 - \rho_0
\label{eq:numer}
\end{equation}
or
\begin{equation}
N_e + \rho_0 = (W_e + 1) Q_0, \qquad W_e = W_0 - e.
\label{eq:numeq}
\end{equation}
Now the exact quota, if recalculated from $N_e$ and $W_e$ for the next stage of counting, is from (\ref{eq:quota})
\begin{equation}
Q_e = \frac{N_e + \rho_e}{W_e + 1}, \qquad \rho_e \in [1, W_e + 1].
\label{eq:quotae}
\end{equation}
Comparing (\ref{eq:quotae}) with (\ref{eq:numeq}), one immediately sees that \emph{if} $\rho_0 \leq W_e + 1$, then $Q_e$ and $Q_0$ are in fact identical.
But if $\rho_0 > W_e + 1$, then necessarily $\rho_e < \rho_0$ and $Q_e$ will be less than $Q_0$ by a certain whole number of votes.
Once again, an exact division formula (of slightly modified type) makes this precise.
Keeping in mind that $\rho_0 \geq 1$, write
\begin{equation}
\rho_0 = q (W_e + 1) + \rho_e, \qquad \rho_e \in [1, W_e + 1],
\label{eq:rhoe}
\end{equation}
and observe that this formula uniquely defines the quotient $q$ and remainder $\rho_e$.
It differs from the ordinary division formula for $\rho_0 / (W_e + 1)$ only in that if $\rho_0$ is an exact multiple of $W_e + 1$ then one additive factor $W_e + 1$ is separated out as the remainder $\rho_e$ rather than leaving it as zero.

Using (\ref{eq:rhoe}) for $\rho_0$ in (\ref{eq:numeq}) and comparing with (\ref{eq:quotae}), it is now apparent that
\begin{equation}
Q_e = Q_0 - q \qquad \text{where} \qquad 0 \leq q \leq \floor{\frac{W_0}{2}}.
\label{eq:quotaeq}
\end{equation}
That is to say, if the quota $Q$ is recalculated after electing $e < W_0$ winners, then $Q$ will reduce by at most a number of votes equal to half the total number of winners, rounded down---this upper bound occurring when $W_e = 1$ ($e = W_0 - 1$) and either $\rho_0 = W_0 + 1$, or $\rho_0 = W_0$ and $W_0$ is odd.
Meanwhile, the lower bound $q = 0$ occurs when $\rho_0 \leq W_e + 1$.
In the happy situation where $\rho_0 = 1$ or $\rho_0 = 2$, one finds that $\rho_e = \rho_0$ and hence $Q_e = Q_0$ irrrespective of the value of $e \in [1, W_0 - 1]$, a situation one may describe as \emph{robust quota}.

The overall outcome may be stated as a formal proposition.
\begin{proposition}[Stability of STV quota]
\label{prop:stvquota}
Suppose that at some stage of an STV count there are $N_0$ votes and $W_0$ winners yet to be elected, of which $e < W_0$ are in fact elected, receiving at least the quota $Q_0$ of votes.
Suppose also that no ballots exhaust when surplus votes are transferred.
Then if the quota $Q$ is recalculated for the next stage, it will reduce by at most $\floor{W_0 / 2}$ votes from $Q_0$, and will never increase from $Q_0$.
Further, if $N_0 \equiv -1$ or $N_0 \equiv -2 \mod (W_0 + 1)$, the recalculated quota will not vary from $Q_0$ for the rest of the count (`robust quota') provided no votes exhaust.
\end{proposition}
This latter case where the notional value of $Q$ does not change at all during the count is illustrated in Table \ref{tab:robustquota}, which illustrates successive stages of an STV election with $W = 6$ winners and $N = 50,000$ votes.
\begin{table}
\begin{centre}
\begin{tabular}{c|c|c|c}
Number elected & Remaining winners & Remaining votes & Quota \\
$e$ & $W_e = W - e$ & $N_e = N - eQ$ & $\floor{N_e / (W_e + 1)} + 1$ \\ \hline
0 & 6 & 50,000 & 7,143 \\
1 & 5 & 42,857 & 7,143 \\
2 & 4 & 35,714 & 7,143 \\
3 & 3 & 28,571 & 7,143 \\
4 & 2 & 21,428 & 7,143 \\
5 & 1 & 14,285 & 7,143
\end{tabular}
\end{centre}
\caption{STV election with 50,000 voters and six winners, illustrating the robust-quota property.}
\label{tab:robustquota}
\end{table}
One has in this case $50,000 \equiv -1 \mod 7$, that is, 50,000 is one less than an exact multiple of $W + 1 = 7$.
Explicit calculation as in Table \ref{tab:robustquota} shows that the integer quota $Q$ then comes out the same (equal to $(N + 1) / (W + 1)$) no matter how many winners $e$ are elected at any given stage.

The robust-quota case, when it occurs, has the additional feature that election outcomes with a continually recalculated quota do not depend at all on whether candidates achieving the quota $Q$ at the same stage are elected \emph{simultaneously} or \emph{sequentially}.
In other cases, there may be slight differences because (for example) simultaneous election of two winners will reduce the effective count by $2Q_0$, while sequential election reduces it by $Q_0 + Q_1$, which may differ by one vote.
This can only affect the ultimate result in improbably close elections, but it underlines the importance of having clear rules to cover the case where two or more candidates achieve a quota simultaneously.

In practice in Australian STV elections, the notional erosion of the quota $Q$ due to exhausted ballots will far outweigh these small variations, were $Q$ to be recalculated after candidates are elected.
(Recalculating $Q$ to take into account exhausted ballots gives what is known as Meek's method in the literature.)
Under current rules for Australian Senate elections, however, the quota is not recalculated even for exhausted votes.
Instead, there are saving provisions stating that if no remaining candidate has a quota, and successive elimination reduces the number of contesting candidates to be equal to the number $W_e$ remaining to be elected, those $W_e$ candidates are elected.
(If $W_e = 1$ then it suffices for one candidate to achieve an absolute majority of remaining votes.)

\subsection{Primary votes, secondary votes and dominance relations}
\label{sec:dominance}

A theme that will come out of more detailed consideration of transferable voting is that it relies on \emph{mixed criteria} to determine an outcome.
That is to say, while it implicitly gives weight to intuitive desiderata that characterise desirable candidates---such as beating other candidates on first-place votes or being ranked above other candidates in one-on-one contests---with perhaps one or two exceptions it does not rely \emph{exclusively} on any one criterion to the exclusion of others.

The first stage of counting in transferable voting considers only each voter's top-ranked candidate.
This top ranking is called the \emph{primary vote}, and is of course the only vote that counts in a FPTP election.
The primary vote for a candidate, in turn, is the number of ballots that rank that candidate in first place.

The one decisive criterion that applies at this first stage, to the exclusion of all other criteria, is the following, which is generally considered an intuitively natural application of the principle of `majority rule':
\begin{proposition}[Quota of primary votes wins]
Any candidate whose primary vote meets or exceeds the quota $Q$ is automatically elected under a transferable voting system.
In an election with only one winner, a candidate with an absolute majority of primary votes is automatically elected.
\end{proposition}
While this property is essentially an automatic consequence of the operation of the system, it should be noted that even this basic property is not satisfied by some voting systems that have been advanced based on other considerations.
(Positional or score-based systems, such as Borda count and its generalisations, furnish many examples.)

Alongside primary votes one may posit a more informal concept of \emph{secondary votes} in transferable voting systems.
These are votes that flow via transfer of preferences from primary votes for `uncompetitive' candidates.
With transferable voting it is common for minor party or independent candidates to stand for election but receive primary votes that are too small to prevent elimination at early stages of the count.
Under FPTP and similar systems such candidates would be advised to withdraw their candidacy as they would otherwise act as `spoilers' drawing votes away from compromise candidates with broader support.
But with transferable voting, such candidates have the opportunity to build up a constituency over several election cycles and demonstrate increasing voter support, so as to eventually occupy a competitive position.
Examples of this are readily identified from voting trends over time for Green and other minor party candidates in Australian elections.

Accordingly, the usual situation in AV elections, and in the contest for the final position in STV elections, is that of a contest between two, three or four front-runners based on combined primary and secondary votes---the latter being transferred from other eliminated candidates.

If this final contest for election is between just two candidates, the outcome will be decided by the familiar tallying of \emph{ranking dominance}, or as Australians would generally say, a \emph{two candidate preferred} count.
Given two candidates $X$ and $Y$, one says that $X$ \emph{dominates} $Y$ or $X$ \emph{is preferred to} $Y$, conventionally written $X \pref Y$, if the number of ballots that rank $X$ above $Y$ is greater than the number that rank $Y$ above $X$.
(When voters are not required to rank all candidates, ballots that rank only one of $X$ or $Y$ count in favour of that candidate, while those that rank neither candidate do not count at all.)
Equivalently, $X \pref Y$ if $X$ would win in a two-candidate runoff against $Y$.
For any pair $X$ and $Y$ it will be the case either that $X \pref Y$, or that $Y \pref X$, or that their votes are tied.
Other than in the rare latter case, this determines the result of the election when there are only two candidates remaining.

The dominance relation among candidates is closely related to the concepts of \emph{Condorcet winner} and \emph{Condorcet loser} that are fundamental in social choice theory.
A Condorcet winner is a candidate $X$ such that $X \pref Y$ for \emph{all} other candidates $Y$, while a Condorcet loser is a candidate $Y$ such that $X \pref Y$ for all other candidates $X$.
Clearly there can be at most one Condorcet winner or Condorcet loser in any election.
It may also be that there is no Condorcet winner or loser, such as when there is a tie for most preferred or least preferred candidate, or when a \emph{Condorcet cycle} exists: a trio of candidates $X$, $Y$, $Z$ such that $X \pref Y$, $Y \pref Z$ and $Z \pref X$.
In large electorates with realistically diverse preferences, a Condorcet cycle can be substantially more likely than a tie.

A familiar criticism of AV for single-winner elections is that while it ensures a Condorcet loser is never elected, it does not guarantee that a Condorcet winner is always elected if one exists.
Note however that a candidate with an absolute majority of primary votes is automatically a Condorcet winner, and in this case \emph{is} guaranteed to be elected under any transferable voting system.

\subsection{Ballot types and combinations}
\label{sec:profiles}

The concepts of \emph{ballot type} and \emph{voting profile} are useful when reasoning about election outcomes.
In a transferable voting system, a \emph{ballot type} for an election with $K$ candidates is simply a ranking of the $K$ candidates, or, where incomplete rankings are allowed, a ranking of some subset of $k \leq K$ candidates together with $K - k$ `nulls' representing the unranked candidates.

The principle underlying the use of ballot types is that in an actual election, two voters casting ballots of the same type are completely interchangeable.
So to describe the full set of votes in an election, it suffices in place of enumerating all ballots individually to simply list all ballot types observed and the number of individual ballots of each type.

However, even an exhaustive list of ballot types can be prohibitive in elections with more than about four candidates.
When voters are required to rank all $K$ candidates there are a total of $K!$ ballot types, corresponding to the full set of permutations of $K$ objects.
This is already a number that expands super-exponentially with $K$, but if voters have the choice not to rank all candidates the number expands still further.
Fortunately, ballots with $K - 1$ candidates ranked and the last spot left blank are equivalent to those with a full ranking, given the $K$th candidate is the only one that can occupy that last spot.
Usually a ballot is required to rank at least some number $K_0$ of candidates to be valid, so the total possible number of ballot types in an election with optional ranking is
\begin{equation}
\text{Possible ballot types} = \typ[K_0]{K} = K! + \sum_{k = K_0}^{K - 2} k! \leq K^2 \cdot (K - 2)!.
\label{eq:numtypes}
\end{equation}
Table \ref{tab:types3} illustrates the $\typ{3} = 9$ possible ballot types for an election with $K = 3$ candidates in which voters may stop after ranking $K_0 = 1$ candidate.
\begin{table}
\begin{centre}
\begin{tabular}{r|ccccccccc}
Type label & $a$ & $\alpha$ & $\bar{a}$ & $b$ & $\beta$ & $\bar{b}$ & $c$ & $\gamma$ & $\bar{c}$
\\ \hline
1st place & A & A & A & B & B & B & C & C & C \\
2nd place & B & C & -- & C & A & -- & A & B & -- \\
3rd place & C & B & -- & A & C & -- & B & A & -- \\
\end{tabular}
\end{centre}
\caption{The nine ballot types with $K = 3$ candidates $A$, $B$, $C$ and allowing partial ranking.}
\label{tab:types3}
\end{table}
If voters are required to rank all candidates, then the three types $\bar{a}$, $\bar{b}$, $\bar{c}$ are invalid and there remain $\typ[3]{3} = 3! = 6$ ballot types.
For larger numbers of candidates $K = 4$ and $K = 5$ the number of possible types expands to $\typ{4} = 27$ and $\typ{5} = 129$ respectively, assuming $K_0 = 1$.

In an election with $N$ voters, each individual voter can be thought of as choosing one of the $\typ[K_0]{K}$ possible ballots when casting their vote, and if they do this independently and with equal likelihood of choosing any one ballot type, it is not difficult to see that the total number of equiprobable \emph{voting profiles} is
\begin{equation}
\text{Voting profiles} = \paren{\typ[K_0]{K}}^N \qquad \text{with $N$ voters and $K$ candidates.}
\label{eq:profiles}
\end{equation}
Each of these voting profiles can be expressed as a `word' made up of $N$ `letters' each of which is a label associated to a particular ballot type.
Thus with $N = 3$ voters, the word $\alpha\alpha\alpha$ means all three vote according to the ranking $A > C > B$, while $abc$ means they vote the different rankings $A > B > C$, $B > C > A$, $C > A > B$.
(The latter is an example of a Condorcet cycle.)

A further equivalence comes about when considering individual voting profiles given that elections are \emph{anonymous}, meaning it is of no importance \emph{which} voters return particular ballots, only the total number of ballots of each type returned.
So for example, the six words $abc$, $acb$, $bca$, $bac$, $cab$, $cba$ representing six distinct voting profiles for three voters do not actually need to be distinguished in practice, and the single `profile type' $abc$ can be used to represent all six permutations.
This profile type effectively counts one vote apiece against each of the three ballot types $a$, $b$, $c$.

Thus, if one is considering the space of all possible voting profiles in an election with $N$ voters it suffices to consider, in place of the full `dictionary' of $\typ[K_0]{K}^N$ words of $N$ letters representing voting profiles, only the slightly smaller number of `ordered words' that have the $N$ letters in some predefined ascending order.
Their number is equivalent to the number of ways of placing $N$ identical objects into $\typ[K_0]{K}$ non-identical boxes (where boxes may be left empty), which is
\begin{equation}
\text{Voting profile types} = \binom{N + \typ[K_0]{K} - 1}{\typ[K_0]{K} - 1} \qquad
\text{with $N$ voters and $K$ candidates.}
\label{eq:ordprofiles}
\end{equation}
For example, with $\typ{3} = 9$ ballot types and $N = 3$ voters, the total number of voting profiles by (\ref{eq:profiles}) is $9^3 = 729$, but when these are arranged into equivalent profile types, the number of distinct types as given by (\ref{eq:ordprofiles}) is just $\binom{11}{8} = 165$.

It must be stressed that if the original profiles enumerated by (\ref{eq:profiles}) are considered equiprobable (`impartial culture' in social choice literature), this gives different probabilities to the profile types enumerated by (\ref{eq:ordprofiles}).
Equiprobability in the latter is also called `impartial anonymous culture'.

\subsection{Tied votes}
\label{sec:ties}

With transferable voting, as in almost any voting system, there is the theoretical possibility of tied votes leading to undetermined outcomes.
This arises in the elimination stage of the count, in the unfortunate case where there are two or more candidates tied for last place but their simultaneous elimination causes the total of elected and contesting candidates remaining to fall below $W$.

In the simplest case of AV elections, or the election of the last winner in STV, a tied result occurs if and only if at some stage of the count there are $k \geq 2$ contesting candidates each with exactly $N/k$ votes (and the total number $N$ of votes is an exact multiple of $k$).
For an STV election where there are two or more winners yet to be elected there is a greater range of possibilities, starting with the scenario where a single candidate falls just short of a quota $Q$ of votes while the remaining candidates are all tied on some lower number, such that the total $N$ of vote values (plus exhausted votes, if $Q$ is not recalculated accordingly) is still consistent with the definition of $Q$ from $N$.

It nonetheless remains true that the probability of any such tie converges to zero as the number $N$ of votes cast increases.
This is a property known as \emph{resolvability}, which transferable voting shares with most other practical systems.
As ties are so rare and reveal a radical absence of social consensus, the rules for breaking ties can be acceptable even if they have a degree of arbitrariness.
Australian election rules generally stipulate that ties in the first instance are resolved by a \emph{countback}: essentially reverting to the most recent stage of the count where the tied candidates have different numbers of votes and then electing the one with the greatest number of votes at that earlier stage.
If a countback fails due to no earlier stage breaking the tie, the winner is determined by lot.

Having noted the above, cases with undetermined outcomes due to ties may be excluded from consideration in what follows.
Nevertheless, as will be seen later, notional tied scenarios can still serve a theoretical purpose as `edge cases' between two separate domains of outcomes.

\section{AV and Dominant Pairs}
\label{sec:pairs}

When transferable voting is used to elect a single winner, there are no complications due to vote transfers from other winners, and the process is purely one of successive elimination.
In practice the process can stop once one candidate attains an absolute majority, but in principle there is nothing that prevents it being carried all the way through until every candidate but the winner is eliminated.
(Stages after an absolute majority is attained may increase the winner's majority but will not change the outcome.)
The only way this process may be blocked from proceeding to this final endpoint is with a tied outcome, where $k \geq 2$ contesting candidates have exactly $N/k$ votes apiece (and none has an absolute majority by definition).

Consideration of the final stage of this full elimination process leads to the following key principle:
\begin{proposition}[Dominant-pair principle for AV]
\label{prop:dompair}
In any AV election there exist two candidates $A$ and $B$ such that the election outcome is determined by the dominance relation (`runoff') between them (in the sense that $A$ wins if $A \pref B$, $B$ wins if $B \pref A$, otherwise there is a tie) and such that every other candidate is eventually eliminated in the count in favour of $A$ or $B$, or ties with them.
Provided no ties occur during the count, the pair $(A,B)$ is uniquely determined and is termed the \emph{dominant pair}; it is made up of the unique winner and a unique \emph{runner-up}.
\end{proposition}
This principle formalises the concept, familiar to Australians, of a \emph{two candidate preferred} or \emph{two party preferred} result, which is simply a binary tallying of votes in favour of $A$ and $B$ within the dominant pair $(A,B)$.
It also justifies the alternative name \emph{Instant Runoff Voting} for AV.
Ties can interfere with the uniqueness of the dominant pair in two distinct ways: if the final outcome is a tie between $k > 2$ candidates then any two of these form a pair satisfying Proposition \ref{prop:dompair}; while if there is a unique winner but $k > 1$ candidates are tied for second place at the final stage, then any one of those $k$ can be paired with the winner to satisfy Proposition \ref{prop:dompair}, meaning the runner-up is not unique.

Many important characteristics of AV follow as immediate consequences of the dominant-pair principle.
Most fundamentally, AV can never elect a Condorcet loser since the winner of an AV election must also be preferred in a runoff against at least one other candidate.
Another key consequence is that except where a tie occurs, if $C$ is any candidate outside the dominant pair $(A,B)$ then it must be the case in at least one stage of the count that both $A$ and $B$ individually have more votes than $C$.
One also has the following:
\begin{proposition}[Irrelevance of dominant-pair preferences in AV]
\label{prop:dompref}
In an AV election where $(A,B)$ is a dominant pair of candidates, the ranking of candidates placed below either $A$ or $B$ in any voter's ballot plays no role in the outcome.
Equivalently, the election result is unchanged if all ballot rankings are considered to be truncated at the point where either $A$ or $B$ appear.
\end{proposition}
This fact about AV ballots will turn out to underlie many of the aspects of AV that are considered paradoxical or problematic.
Note that Proposition \ref{prop:dompref} should be considered to exclude tied scenarios involving more than two candidates: election rules in this situation may call for a countback to eliminate one of the tied candidates, meaning that the preferences of this candidate do count.

An important practical property of AV elections is that regardless of the number of candidates and the exponential possibilities raised for the number of ballot types, it is nonetheless possible in the majority of practical cases to identify the dominant pair, or at least a smaller set of candidates guaranteed to contain the dominant pair, by inspection of the primary votes alone for each candidate.
\begin{proposition}[Dominant pair from primary vote in AV]
\label{prop:domprim}
If in an AV election there are two candidates $A$ and $B$, each of whose individual share of primary votes is greater than the total share of primary votes for all remaining candidates, then $(A,B)$ is a unique dominant pair.

More generally, if there is a subset $S_k$ of $k \geq 2$ candidates, each of whose share of primary votes is greater than the combined share for all candidates not in $S_k$, then any dominant pair will comprise two candidates from $S_k$.
Such a set $S_k$ will be referred to as a \emph{dominant $k$-set}.
\end{proposition}
Proposition \ref{prop:domprim} is evident from the fact that under the stated conditions, candidates in the set $S_k$ will never be eliminated prior to those outside $S_k$, and it is not possible for any candidate outside $S_k$ to achieve a count of votes from transfer of preferences that equals that of any candidate in $S_k$.
This provides a formalisation of the concept of \emph{uncompetitive candidate} noted previously in regard to secondary votes; a candidate is uncompetitive if they fall outside the smallest dominant $k$-set satisfying the conditions of Proposition \ref{prop:domprim}.

A simple algorithm may be formulated for identifying dominant $k$-sets for any given election.
First, place all candidates in order from greatest to least number of primary votes.
Then proceed from bottom to top, keeping a cumulative total of primary votes, and flag each candidate for which the primary vote about to be added to the cumulative total is greater than the total so far.
When the top is reached, the highest flagged candidate together with all candidates above forms the smallest set $S_k$ satisfying Proposition \ref{prop:domprim}.
If this is the first placed candidate, they win by absolute majority; if it is the second placed candidate then the dominant pair has been successfully identified.

(Observe also that if there are candidates with identical numbers of primary votes, then the sets $S_k$ identified by the above procedure will always include either all or none of those candidates, so the order in which those candidates are listed does not matter.)

It may of course transpire that there \emph{is} no dominant $k$-set smaller than the set of all candidates itself (or all candidates bar the one placing last).
This may be termed a \emph{fully competitive} election, and requires a full distribution of preferences to be undertaken before the winner or the dominant pair can be determined.
In a fully competitive election the elimination of the last-placed candidate(s) is still assured, but it is possible for the candidate placing next-to-last on primary votes to accumulate preference flows from other eliminated candidates and ultimately emerge as the winner.
Fully competitive elections are found to be relatively unlikely in practice, however.

In order for all $k$ candidates in a dominant $k$-set to have a number of primary votes greater than the aggregated `residue' vote $R_k$ for candidates outside the set, it is necessary that
\begin{equation}
R_k < \frac{N}{k + 1} \qquad \text{aggregated for all candidates outside a dominant $k$-set.}
\label{eq:residuek}
\end{equation}
A little consideration reveals that $R_k$ is maximised when each candidate within the $k$-set obtains a number of primary votes equal or almost equal to the Droop quota $Q_k$ calculated by (\ref{eq:droop}) based on $k$ `winners'.
(Strictly speaking, because of rounding considerations it is sometimes possible for some candidates in the $k$-set to fall short of $Q_k$ by a small number of individual votes while still exceeding the largest possible value of $R_k$.)
If $R_k$ is less than this maximal value, it is possible for some candidates in the $k$-set to have substantially fewer than $Q_k$ primary votes, while other candidates will have more than $Q_k$ votes to make up the overall tally of $N$ primary votes.
The following will always be true in any case:
\begin{proposition}[Plurality quota for dominant $k$-set in AV]
\label{prop:domquota}
In an AV election, a necessary condition for existence of a dominant $k$-set in accordance with Proposition \ref{prop:domprim} for any $k \geq 2$ is that at least one candidate receives $Q_k = \floor{N / (k + 1)} + 1$ or more primary votes, equivalent to the Droop quota for $k$ winners in an STV election.
\end{proposition}
The special case $k = 2$ is that where the dominant pair for the AV election can be identified at once from the primary votes alone.
This requires at least one candidate (say $A$) to obtain a minimum of $Q_2 = \floor{N / 3} + 1$ primary votes.
Denoting by $N_A$ the primary vote for $A$, there is then a second candidate (say $B$) with $N_B \leq N_A$ votes, with all other primary votes adding to a residue $R_2 < N_B$.

Since $N_A + N_B + R_2 = N$, it follows that if $N_A = N_B = Q_2$ then $R_2 = N - 2 Q_2$ and this is guaranteed to be strictly less than $Q_2$.
Using the exact formula (\ref{eq:quota}) for the Droop quota it will be found that in fact $R_2 = (N - 2 \rho) / 3$, where the modified remainder $\rho$ is either 1, 2 or 3, and it is sometimes possible to allow $R_2$ to increase by one vote while maintaining $N_B > R_2$.
Specifically:
\begin{itemise}
\item
if $N \equiv 2 \mod 3$ ($\rho = 1$) then the maximum possible value of $R_2$ is $\floor{N / 3} = (N - 2) / 3$ and requires that $N_A = N_B = Q_2 = (N + 1) / 3$;
\item
if $N \equiv 1 \mod 3$ ($\rho = 2$) then the maximum possible value of $R_2$ is $\floor{N / 3} - 1 = (N - 4) / 3$ and requires that $N_A = N_B = Q_2 = (N + 2) / 3$; and
\item
if $N$ is an exact multiple of 3 ($\rho = 3$) then the maximum possible value of $R_2$ is $N / 3 - 1$ and requires that $N_A = Q_2 = N / 3 + 1$ and $N_B = Q_2 - 1 = N / 3$.
\end{itemise}
Note in particular that in the second case above $N_B$ exceeds $R_2$ by two votes, but it is not possible to hypothetically increase $R_2$ by one vote to $\floor{N / 3}$ votes without reducing $N_B$ to also become equal to $\floor{N / 3}$ votes.
It is mathematically (if not practically) plausible in that situation that preferences could flow entirely within the `residual' group as candidates are eliminated, so that some candidate $C$ ultimately ties with $B$ on $\floor{N / 3}$ votes, leaving $A$ with $\floor{N / 3} + 1$ votes.
Were this rather contrived situation to occur in reality, the rules would determine either that $B$ and $C$ are eliminated simultaneously leading to the election of $A$, or that a countback finds an earlier stage where $C$ had fewer votes than $B$ and will accordingly be eliminated leading to a runoff between $A$ and $B$ after distributing $C$'s preferences.
Australian election rules generally call for the latter, meaning that \emph{in practice}, $(A,B)$ can be considered a dominant pair even in this situation.

Remaining with the case $k = 2$, if candidate $A$ receives $N_A > Q_2$ primary votes, then a candidate $B$ with $N_B < Q_2$ may still form a dominant pair with $A$ provided
\begin{equation}
N_B > R_2 = N - N_A - N_B, \qquad \text{or} \qquad N_B > \frac{N - N_A}{2}.
\label{eq:dompair}
\end{equation}
This of course is of practical importance only if $N_A$ falls short of an absolute majority, in which case candidate $B$ needs more than $N / 4$ primary votes to be assured of a runoff with $A$.

Accordingly the full conditions for the existence of a dominant 2-set without an absolute majority are a first-placed candidate $A$ with between one-third and one-half of the primary vote, and a second-placed candidate $B$ with at least one-quarter of the primary vote, such that the total of all remaining primary votes is less than the number of votes for $B$.

A related fact about AV elections follows readily by considering the flow of preferences within subsets of candidates:
\begin{proposition}[Stable coalitions win in AV]
\label{prop:avcoalition}
In an AV election, if there exists a subset (`coalition') $C$ of candidates such that an absolute majority of voters rank all candidates in $C$ ahead of any candidate outside $C$, then the winner will also come from the set $C$.
However, it is possible that neither the dominant pair nor any dominant $k$-set for any $k \geq 2$ includes candidates from $C$ alone.
\end{proposition}
This proposition formalises the practice of `swapping preferences' which is also common in Australian elections.
In Senate elections prior to 2016 it was common practice for small parties to submit party lists that directed preferences to other small parties ahead of any other candidates, which often ensured that the contest for the last winner (essentially an AV election) became a \emph{de facto} lottery among minor-party candidates.
In the current Senate and most other Australian elections, such stable coalitions can only emerge from voters directing their own preferences, which many argue is a much more transparent arrangement and one that more consistently respects voter choice.

A simple example illustrates the second part of Proposition \ref{prop:avcoalition}.
Suppose there is one candidate $A$ with 45\% of primary votes, another candidate $B$ with 30\% of primary votes, and $C$ is the set of all remaining candidates who collectively tally 25\% of primary votes.
Suppose further that all voters who put $B$ or any of the candidates in $C$ in first place also put $A$ in last place, or equivalently do not rank $A$.
These voters form an absolute majority of 55\% of the electorate and ultimately cause $B$ to be elected over $A$ after distribution of preferences from $C$, with $B$ and $C$ together comprising the stable coalition.
However, the dominant pair in this election is $(A,B)$ and any dominant $k$-set will necessarily contain $A$.

(One may also note that in this example $A$ is a Condorcet loser who would nonetheless be elected under FPTP and potentially under other popular methods such as a Borda count.)

\section{AV and Condorcet Winners}
\label{sec:condorcet}

The fact that there can exist a Condorcet winner in a single-winner election that is not elected by AV is often argued to be a fatal deficiency, contrary to the principle of majority rule.
(See for example \cite{m:uaes}.)
Certainly from the classical point of view where one aims to determine the most suitable social \emph{ranking} from a set of individual voter rankings, the idea that a Condorcet winner cannot reasonably appear anywhere but at the top of the social ranking is almost irresistible.

The aim of AV elections however is not to determine a social ranking of candidates but to distil the will of the electorate as best one can given the winner-take-all nature of the contest.
The implicit aim of AV, as Thomas Hare and John Stuart Mill themselves explained, is to balance the \emph{comparative} assessment of candidates by voters with a concept of a mandate expressed through the placing of a favoured candidate in a high \emph{absolute} position by sufficient voters.

It is helpful to recapitulate some of the necessary conditions for a candidate $C$, who is a Condorcet winner, to \emph{not} be elected by AV:
\begin{itemise}
\item
Candidate $C$ cannot have an absolute majority of primary votes.
\item
Candidate $C$ cannot be one of a dominant pair in the sense of Proposition \ref{prop:dompair}, hence must, at some stage of the count, have fewer votes than either candidate in the dominant pair.
\item
$C$ is nevertheless preferred to both dominant-pair candidates by an absolute majority of voters.
\end{itemise}
These conditions appear difficult to reconcile, but can hold simultaneously in situations where $C$ scores strongly on `second place' votes across a sufficently large number of ballots.
It is not difficult to conceive scenarios involving two strong candidates $A$ and $B$ and a `compromise' candidate $C$ such that most $A$ voters prefer $C$ to $B$, while most $B$ voters prefer $C$ to $A$.
In this situation $C$ would be the clear winner if their \emph{primary} vote support was greater than for at least one of $A$ or $B$, but without such support $C$ can be eliminated before either, leading to a runoff between $A$ and $B$.

Note that this can also extend to the case where $A$ and $B$ are `stable coalitions' in the sense of Proposition \ref{prop:avcoalition}: that is, sets of two or more candidates whom voters rank completely ahead of all other candidates, including $C$.
This can mean that all individual candidates have fewer primary votes than $C$, albeit the aggregate primary vote of each coalition will be greater than for $C$.
It does require that sufficiently many voters rank $C$ in the intermediate position between candidates from $A$ and candidates from $B$ or vice versa.

In the worst case the primary vote may split almost equally for each candidate or coalition in a three-way contest, with $C$ receiving just below one-third of primary votes and being eliminated.
That this is the worst case is the substance of the following proposition:
\begin{proposition}[Condorcet winner in AV]
\label{prop:condorcet}
In an AV election with $N$ formal ballots, a Condorcet winner who has at least $Q_2 = \floor{N / 3} + 1$ primary votes, or who subsequently gains this many votes by transfer of preferences while still contesting, is always elected.
\end{proposition}
To verify Proposition \ref{prop:condorcet}, note first that for a Condorcet winner to be elected it is sufficient that they be one of a dominant pair.
Therefore, assuming Condorcet winner $C$ has $N_C \geq Q_2$ votes, by Proposition \ref{prop:domprim} it suffices that there is some other candidate $B$ with more than $(N - N_C) / 2$ primary votes.
If this is not the case, let the count proceed by successive elimination of candidates and note that $C$, having $Q_2$ or more votes, will not be eliminated provided at least two other contesting candidates remain, nor will $C$'s vote decrease.
It is also not possible for some other candidate, say $A$, to accumulate an absolute majority by transfer of preferences while $C$ remains contesting since this would imply that $A \pref C$.
Thus either $C$ eventually accumulates an absolute majority and is elected, or else the count reaches a stage with three contesting candidates $A$, $B$, $C$, none with an absolute majority, in which case one out of $A$ or $B$ must have strictly fewer votes than $C$ and be eliminated causing $C$ to be elected in a runoff against the other.

All these considerations help to firm up the circumstances where a Condorcet winner $C$ fails to be elected under AV.
It is now clear that $C$ must receive fewer than one-third of primary votes and must fail to reach a threshold of one-third of votes through transfer of preferences, before being eliminated in favour of two strong candidates or coalitions with higher vote tallies.
Their status as a Condorcet winner also depends on them receiving sufficient second or lower-placed preferences from those stronger candidates or coalitions, that they would technically beat each one individually in a runoff.
$C$ could be seen as a victim of Proposition \ref{prop:dompref} and the fact that their tendency to rank below any one dominant-pair candidate, but above the other, is not able to influence the result.

Note that this scenario in no way resembles the kind of bad outcome seen in FPTP, where a candidate who might command an absolute majority is excluded due to the presence of `spoiler' candidates that appeal to some of the same political constituencies.
This is instead the situation of one who has broad appeal as a `compromise' candidate but is `beaten into third place' by other candidates that rank more highly on more individual voters' ballots.
This, it may be argued, is less the travesty of democracy it is made out to be, and more an acknowledgement that AV is not a pure ranking system but instead gives \emph{joint} recognition to relative considerations (ranking of preferences) and absolute considerations (placing first on enough individual ballots).

It has been pointed out in the social choice literature \cite{rg:pppidav} that it is quite possible for a candidate to be a Condorcet winner despite receiving \emph{no} primary votes whatsoever.
This does require a minimum of four candidates, since with only three candidates where one receives no primary votes, one other candidate would have an absolute majority or there would be a tied outcome.
But with four candidates $(A,B,C,D)$ it is easy to postulate a scenario where $A$ is a single candidate with plurality support, $(B,D)$ form a preference-swapping coalition against $A$, and $C$ is a compromise candidate that all voters rank second.
Suppose then that 40\% of ballots follow the profile $A > C > B > D$, another 35\% follow the profile $B > C > D > A$ and the remaining 25\% follow the profile $D > C > B > A$.
One may readily verify that $C$ is a Condorcet winner despite not appearing first on anyone's ballot.
The AV result is that $C$ and $D$ are eliminated and $B$ wins the runoff against $A$ with $D$'s preferences, as though $C$ were not a candidate.

This example once again serves as a springboard for philosophical dispute, but it is at least defensible (and not necessarily inconsistent with `majority rule') to argue that a candidate not ranked first by a single voter should be passed over for election despite having broad support in purely relative terms.
One may consider that, were this instead a multi-winner election governed by the principle of proportional representation, there would be no paradox involved in electing $A$, $B$ and $D$ ahead of $C$.
Note too that $C$ would no longer be the Concorcet winner were $D$'s voters (or at least 60\% of them) to prefer coalition partner $B$ to the compromise candidate $C$.
Were this the case, the election would proceed exactly as before but now $B$ would be both the Condorcet winner and the successfully elected candidate, in accordance with Proposition \ref{prop:condorcet}.

This section concludes with one further case study, motivated by Proposition \ref{prop:condorcet} and its `one third' threshold which highlights that three-way contests are closely connected with paradoxes around Condorcet winners and AV.
Here a `toy' scenario is formulated and used as the basis for a model of certain constrained types of three-way contest.

Consider a pared-down `election' scenario where there are three candidates $A$, $B$, $C$ and just three voters.
Voters are assumed to rank all three candidates, so the ballot types available to them are those labelled $a$, $\alpha$, $b$, $\beta$, $c$, $\gamma$ in Table \ref{tab:types3}.
The three types labelled with Roman letters are cyclic permutations of the ranking $A > B > C$, while those labelled with Greek letters do similar for the ranking $A > C > B$.

As described following Table \ref{tab:types3}, each possible voting profile is expressible as a three-letter word drawn (with replacement) from the letters $(a,b,c,\alpha,\beta,\gamma)$.
Examples of these include $aaa$ and $b\beta\gamma$.
There are $6^3 = 216$ such profiles, but when these are placed in `dictionary order' recognising that the voters themselves are interchangeable, their number reduces to $\binom{8}{5} = 56$ profile types.

Of the 56 profile types, two are special: $abc$ and $\alpha\beta\gamma$ result in Condorcet cycles, so fail to yield a Condorcet winner.
(These are the classic \emph{Condorcet profiles} originally recognised by their namesake in the 18th century.)
Each of these corresponds to six words in the original unordered voting profiles, hence of the 216 original profiles 12 (one in 18) fail to have a Condorcet winner.
In every one of the remaining profiles a Condorcet winner can be identified:
\begin{itemise}
\item
in 48 out of the 56 ordered profiles (168 out of 216 original profiles, or 7 out of 9) there is one candidate with an absolute majority who is also a Condorcet winner;
\item
while in the remaining 6 out of 56 (36 out of 216, or 1 out of 6), each voter places a different candidate first, but inspection of preferences yields a Condorcet winner regardless.
\end{itemise}
An example of the latter is the profile $ab\gamma$, where all three voters put candidate $B$ in either first or second place and this makes $B$ the Condorcet winner.

This pared-down scenario is not of much value in itself as a model for AV elections, given that the outcome is always either an absolute majority or a tie.
What is more interesting is to consider each `voter' in this model as standing for one-third of a larger voting population, and then considering small changes in the relative sizes of these three proportions.
Tied scenarios, such as for the profile $ab\gamma$, then turn out to be the `edge cases' between two different sets of election outcomes.

It now turns out that in \emph{every one} of these scenarios where the three primary votes all differ, a small perturbation in the relative size of these three voting `blocs' can hand the election to any one of $A$, $B$ or $C$, and can do so regardless of whether one of these is a Condorcet winner, or whether the underlying profile is a Condorcet cycle like $abc$.
Meanwhile, the scenarios with absolute majorities, such as $aab$, behave as expected, where even under fairly large perturbations the majority candidate retains their absolute majority and the Condorcet winner is also the one elected by AV.

This tendency can be found to extend to the case where the three-profile constraint is relaxed and blocs of voters may cast their ballots from all six of the available profiles.
So for example, if one starts from the profile type $aab$ and assumes that a proportion of the majority casting $a$ ballots switch to the alternative profile $\alpha$---leading to a mixture of profiles $aab$, $a\alpha b$ and $\alpha\alpha b$---this does not affect $A$'s absolute majority status.
Meanwhile, if one starts from the Condorcet profile $abc$ and supposes that around half of those voting $c$ switch to profile $\gamma$ instead (swapping their second and third preference), one has a scenario---a fifty-fifty mix of $abc$ and $ab\gamma$---where the very existence of a Condorcet winner (in this case $B$) depends on an exchange of preferences within a relative handful of voters, all of whom have given their primary vote to \emph{another} candidate (in this case $C$).

What these examples suggest is that in many cases where a Condorcet winner lacks an absolute majority of primary votes, or something close to a majority, their status as a Condorcet winner is in a sense \emph{unstable}---a concept that will be revisited after the next section.
Condorcet's original paradox already pointed out more than two centuries ago that a consistent voting system must sometimes elect a candidate $A$ despite an absolute majority of voters preferring some other candidate to $A$.
What AV provides is an assurance that in those situations where $B \pref A$ and yet $A$ is elected, there are grounds for arguing $A$ nonetheless has a majority mandate: in particular there is a guarantee that $A \pref C$ for at least one other candidate $C$ with strong voter support.
The key principle of `majority rule' has not, on this view, actually been abnegated: rather, there is recognition that the principle has various, sometimes conflicting, referents in everyday language, and a satisfactory voting procedure should seek to balance these various senses in which people understand the term.

\section{AV and Non-Monotonicity}
\label{sec:monotone}

\emph{Monotonicity} is the property of voting systems whereby if one or more voters were to alter their individual votes to the advantage (disadvantage) of a particular candidate $X$, this should never change the overall result to the disadvantage (to the advantage) of candidate $X$.

It is a well-known fact that AV is vulnerable to non-monotonicity, which can arise in two ways:
\begin{description}
\item[Winner becomes loser]
In a scenario where candidate $X$ is elected by AV, an alteration that gives $X$ a higher ranking in one or more individual ballots causes $X$ to lose to another candidate $Y$.
\item[Loser becomes winner]
In a scenario where candidate $X$ is not elected by AV, an alteration that gives $X$ a lower ranking in one or more individual ballots causes $X$ to be elected.
\end{description}
The mechanism by which this occurs in both cases is that the election of $X$ relies on a flow of preferences to $X$ from some other candidate $Y$, hence depends on $Y$ being eliminated---in particular, \emph{not} becoming the second of a dominant pair with $X$.
This in turn implies the final runoff is between $X$ and some other candidate $Z$ on whose preferences $X$ does not rely.
In this runoff, monotonicity \emph{does} apply in respect of $Y$ voters: should they favour $X$ more, this does not alter the outcome.

The key here is, rather, that if some of those voting for $Z$ alter \emph{their} vote to prefer $X$ above $Z$, this can reduce $Z$'s vote to fall below that for $Y$ and shift the dominant pair from $(X,Z)$ to $(X,Y)$.
If the preferences of the remaining $Z$ voters are predominantly for $Y$ rather than $X$, this will swing the election in $Y$'s favour.
This hypothetical exchange in votes between $Z$ and $X$ can occur in either direction: to swing the election from $X$ to $Y$ by increasing votes for $X$ (so that $Y$ eliminates $Z$), or to swing it from $Y$ to $X$ by reducing votes for $X$ (so that $Z$ eliminates $Y$).

Once again, the paradox is one that arises in three-way contests where no two candidates have a clear majority over the third and so the dominant pair can be altered by a small swing in votes.
Motivated by this idea, and following the work of Lepelley et al \cite{lcb:lompire}, necessary and sufficient conditions are developed here for AV elections with three candidates to be vulnerable to non-monotonicity, and the consequences of this examined.

\subsection{Winner becomes loser}
\label{sec:winlose}

Return once more to the ballot types set out in Table \ref{tab:types3} for three candidates $A$, $B$, $C$, and assume once again that voters provide a full preference ranking, so restricting to the six types $a$, $b$, $c$ (the cyclic permutations of $A > B > C$) and $\alpha$, $\beta$, $\gamma$ (the cyclic permutations of $A > C > B$).
Assuming $N$ sufficiently large one may set aside considerations due to ties and rounding, focussing purely on the shares of ballots of each type.
With some abuse of notation, the labels $a,b,c,\alpha,\beta,\gamma$ will be used here to refer also to the proportion of total votes cast with each ballot type, such that
\begin{equation}
a + \alpha + b + \beta + c + \gamma = 1
\label{eq:shares3}
\end{equation}
and each voting profile is completely described by the six numerical values in $[0,1]$, subject to (\ref{eq:shares3}).

Without loss of generality, one now seeks conditions such that $A$ is elected by AV from the dominant pair $(A,B)$, and yet a change in voter first preferences from $B$ to $A$ (that is, from profiles $b$ or $\beta$ to $a$ or $\alpha$) switches the dominant pair to $(A,C)$ and causes $C$ to be elected instead.
Observe once again that it suffices to consider only changes in $B$'s primary vote, since any change in votes for $C$ in a scenario where $C$ is eliminated will not change the outcome, nor will $B$ voters' preferences affect the outcome as long as $(A,B)$ remains the dominant pair.

Let $b' \leq b$ and $\beta' \leq \beta$ denote the overall share of votes that swing from $B$ first to $A$ first in such a scenario (irrespective of how the second and third rankings change within these altered ballots) and that the other ballots remain unchanged.
The necessary conditions will then amount to the following:
\begin{itemise}
\item
$a + \alpha < 1/2$, $b + \beta < 1/2$ and $c + \gamma < 1/2$ (no absolute majority in base scenario).
\item
$c + \gamma < \min\paren{a + \alpha, b + \beta}$ ($C$ is eliminated in base scenario).
\item
$a + \alpha + c > 1/2$ and $b + \beta + \gamma < 1/2$ ($A \pref B$ in base scenario).
\item
$a + \alpha + b' + \beta' < 1/2$ ($A$ has no absolute majority after swing).
\item
$(b - b') + (\beta - \beta') < c + \gamma$ ($B$ is eliminated after swing).
\item
$c + \gamma + (b - b') > 1/2$ and $a + \alpha + b' + \beta < 1/2$ ($C \pref A$ after swing).
\end{itemise}
Closer inspection confirms that the preferences of $A$'s voters themselves (that is, the distinction between profiles $a$ and $\alpha$) has no bearing on non-monotonicity for $A$, which allows for reducing the `degrees of freedom' from the six original variables without loss of information.
Given the nature of the conditions it will also prove handy to work with the three variables
\begin{equation}
\rema = \half - \paren{a + \alpha}, \qquad
\remb = \half - \paren{b + \beta}, \qquad
\remc = \half - \paren{c + \gamma},
\label{eq:remabc}
\end{equation}
which represent the shares by which each primary vote falls short of an absolute majority.
By the first condition above each of these `remainders' must be positive, while (\ref{eq:shares3}) becomes
\begin{equation}
\rema + \remb + \remc = \half.
\label{eq:remshares}
\end{equation}
(Note that the case $\rema = \remb = \remc = 1/6$ is the familiar three-way tie scenario.)
The remaining conditions above for non-monotonicity now read as
\begin{gather}
\remc > \max\paren{\rema, \remb}, \label{eq:winlose1} \\
c > \rema \qquad \text{and} \qquad \gamma < \remb, \label{eq:winlose2} \\
\remc - \remb < b' + \beta' < \rema, \label{eq:winlose3} \\
b - b' > \remc \qquad \text{and} \qquad \beta + b' < \rema. \label{eq:winlose4}
\end{gather}
It may be observed that of these four conditions, the first (\ref{eq:winlose1}) deals only with the candidate's primary votes, the second (\ref{eq:winlose2}) with the disposition of $C$'s preferences and the last two with the disposition of $B$'s preferences.
Condition (\ref{eq:winlose1}) when taken in conjunction with (\ref{eq:remshares}) implies that there is an upper bound of $1/4$ on $\rema$ and $\remb$ separately, and a \emph{lower} bound of $1/6$ on $\remc$.
There is accordingly an upper bound of $1/3$ on $\rema + \remb$, which by definition of $\remc$ is also equal to $c + \gamma$.
However, condition (\ref{eq:winlose3}) also implies that $\remc < \rema + \remb = 1/2 - \remc$, which imposes the stronger condition that
\begin{equation}
\frac{1}{6} < \remc < \frac{1}{4} \qquad \text{and} \qquad
\frac{1}{4} < \rema + \remb = c + \gamma < \frac{1}{3}.
\label{eq:nonmonrem}
\end{equation}
This expresses mathematically the fact that if $\remc$ is too small ($C$'s primary vote is too high) then $(A,B)$ cannot be the dominant pair in the base scenario, while if $\remc$ is too large ($C$'s primary vote is too low) then $C$ cannot be promoted into a dominant pair by reducing $B$'s vote.

Turning to condition (\ref{eq:winlose4}), this asserts that
\begin{equation}
b' < \min\paren{b - \remc, \rema - \beta},
\label{eq:winlose4a}
\end{equation}
and since $b'$ cannot be negative this in turn implies that
\begin{equation}
b > \remc \qquad \text{and} \qquad \beta < \rema.
\label{eq:winlose4b}
\end{equation}
In fact, as a consequence of (\ref{eq:remabc}) and (\ref{eq:remshares}) one has $\rema + \remc = b + \beta$ and hence $b - \remc = \rema - \beta$ as an identity, so it suffices to ensure that one or other is positive.
Conditions (\ref{eq:winlose4b}) and (\ref{eq:winlose2}), on the preference shares within $B$'s and $C$'s primary votes respectively, are complementary and ensure the preferences of the eliminated candidate elect $A$ in the base scenario and $C$ in the modified scenario.
One may verify that they also ensure that $A \pref B$ and $C \pref A$ in \emph{either} scenario.
To sum up:
\begin{proposition}[Winner becomes loser in AV]
\label{prop:winlose}
Given an AV election having three candidates $(A,B,C)$, a voter profile with $A$ the winner and $B$ the runner-up is vulnerable to non-monotonicity of the `winner becomes loser' type under the following necessary and sufficient conditions.
\begin{enumerate}
\item
There exist numbers $\rema$, $\remb \in (0,1/4)$ with $\rema + \remb > 1/4$, $\rema + 2\remb < 1/2$ and $2\rema + \remb < 1/2$ such that the primary vote shares for $(A,B,C)$ are $1/2 - \rema$, $1/2 - \remb$ and $\rema + \remb$, respectively.
In particular, no candidate has a majority, but all have more than $1/4$ of votes.
\item
The share $\beta$ of all voters selecting the profile $B > A > C$ is less than $\rema$.
\item
The share $\gamma$ of all voters selecting the profile $C > B > A$ is less than $\remb$.
\end{enumerate}
Then for any choice of ballot shares $b' \in [0, \rema - \beta)$ and $\beta' \in [0, \beta]$ such that
\begin{equation}
b' + \beta' > \half - \rema - 2\remb > 0,
\label{eq:winlose}
\end{equation}
the notional transfer of $b'$ ballots of type $B > C > A$ and $\beta'$ ballots of type $B > A > C$ to give first preference to $A$ instead (with later preferences arbitrary) will cause $C$ instead of $A$ to be elected.

Under the above conditions the set of values $(b',\beta')$ satisfying (\ref{eq:winlose}) is never empty.
Also $A \pref B$ and $C \pref A$ throughout, so either a Condorcet cycle exists or $C$ is the Condorcet winner.
\end{proposition}
Observe in particular that the strictness of the inequality $\rema + \remb > 1/4$ is essential to this result.
(That is to say, all three candidates must receive strictly more than one-quarter of votes.)
Using the fact that there exists $\epsilon > 0$ with $\rema + \remb = 1/4 + \epsilon$, one may always choose $b' = \rema - \beta - \epsilon$ and $\beta' = \beta$ in Proposition \ref{prop:winlose} with (\ref{eq:winlose}) rendered into the identity $\rema - \epsilon > \rema - 2\epsilon$.
The small $\epsilon$ is needed since if one had $b' = \rema - \beta$ exactly, this would also mean $b' = b - \remc$ and with $b - b' = \remc$ this would result in $A$ tying with $C$ instead of losing to $C$ after the swing.

\begin{figure}[t]
\begin{centre}
\begin{picture}(350,350)(-25,-25)
\put(-5,-5){\makebox(0,0)[tr]{0}}
\put(-25,0){\vector(1,0){350}}
\put(325,-10){\makebox(0,0)[t]{$\rema$}}
\put(200,-10){\line(0,1){10}}
\put(200,-15){\makebox(0,0)[t]{$1/6$}}
\put(300,-10){\line(0,1){20}}
\put(300,-15){\makebox(0,0)[t]{$1/4$}}
\put(0,-25){\vector(0,1){350}}
\put(-10,325){\makebox(0,0)[r]{$\remb$}}
\put(-10,200){\line(1,0){10}}
\put(-15,200){\makebox(0,0)[r]{$1/6$}}
\put(-10,300){\line(1,0){20}}
\put(-15,300){\makebox(0,0)[r]{$1/4$}}
\put(0,300){\line(1,-1){300}}
\put(95,135){\rotatebox[origin=t]{-45}{\labelbox{$\rema + \remb = 1/4$\\$(\remc = 1/4)$}}}
\put(0,300){\line(2,-1){200}}
\put(58,260){\rotatebox[origin=b]{-26.57}{\labelbox{$\rema + 2\remb = 1/2$\\$(\remc - \remb = 0)$}}}
\put(200,200){\line(1,-2){100}}
\put(230,110){\rotatebox[origin=b]{-63.43}{\labelbox{$2\rema + \remb = 1/2$\\$(\remc - \rema = 0)$}}}
\put(200,200){\circle*{5}}
\put(205,205){\makebox(0,0)[bl]{\labelbox{Three-way tie\\$\rema = \remb = \remc = 1/6$}}}
\multiput(0,200)(20,0){10}{\line(1,0){10}}
\multiput(200,0)(0,20){10}{\line(0,1){10}}
\end{picture}
\end{centre}
\caption{Visualisation of profiles $(\rema,\remb)$ vulnerable to non-monotonicity under Proposition \ref{prop:winlose}.}
\label{fig:winloseset}
\end{figure}
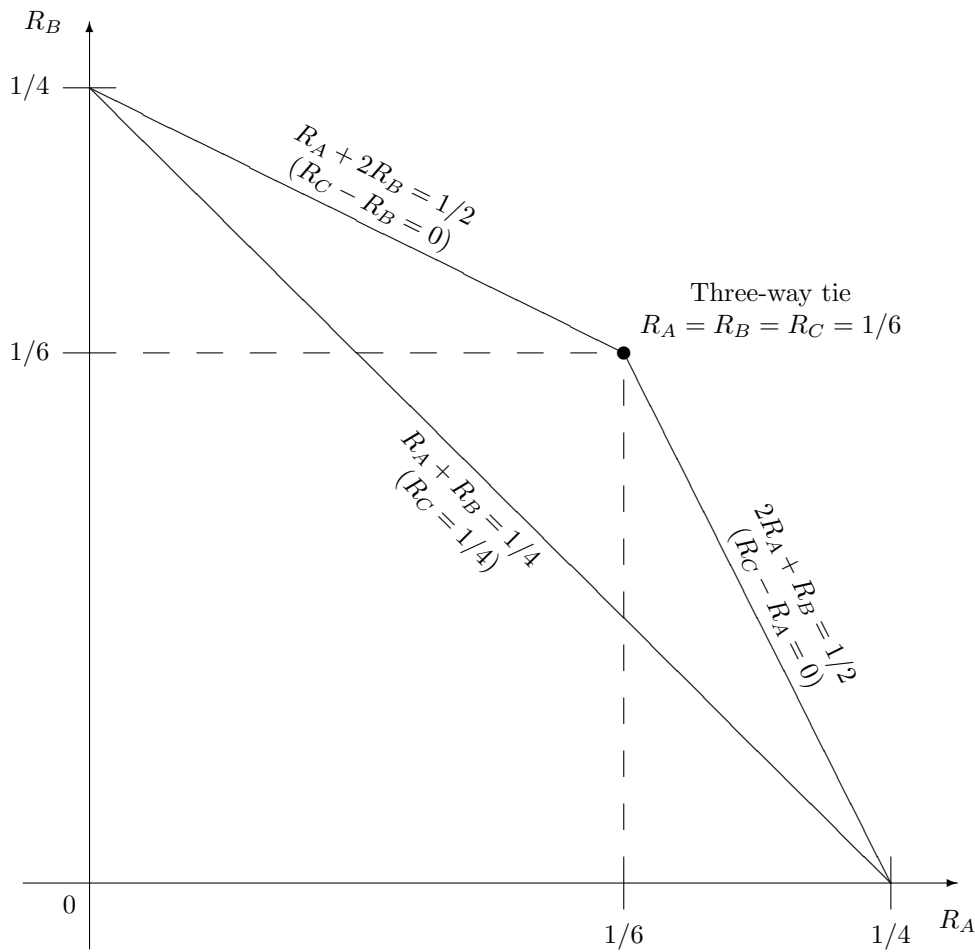

Figure \ref{fig:winloseset} illustrates the set of profiles that are \emph{potentially} vulnerable to `winner becomes loser' under Proposition \ref{prop:winlose} based on the three candidates' primary votes alone, as a point set in the $\rema$--$\remb$ plane.
This point set is the interior of the isosceles triangle defined by the inequalities $\remc < 1/4$ and $\remc > \max(\rema, \remb)$ with $\remc$ implicitly defined by (\ref{eq:remshares}), and thus satisfies condition 1 of Proposition \ref{prop:winlose}.
Voting profiles within this set must additionally satisfy conditions 2 and 3 relating to the second preferences of voters for $B$ and $C$, in order for a latent non-monotonicity to be present.

\subsection{Likelihood of non-monotonicity: a quasi-impartial model}
\label{sec:likelihood}

One may readily calculate by inspection of Figure \ref{fig:winloseset} that the triangle defining the region of potential `winner becomes loser' non-monotonicities has a base length of $\sqrt{2} / 4$ and a height of $(1/6 - 1/8) \sqrt{2} = \sqrt{2} / 24$, hence has area $1 / 96$.
It is tempting to say that this area measure carries some information about the \emph{likelihood} of a non-monotonicity occurring, under suitable conditions.
To make this more precise requires formulating a broader model for three-candidate elections and what values of $\rema$ and $\remb$ (and implicitly $\remc$) can occur more broadly.

\begin{figure}[t]
\begin{centre}
\begin{picture}(350,350)(-170,-170)
\put(-170,0){\vector(1,0){350}}
\put(175,5){\makebox(0,0)[b]{$\rema$}}
\put(0,-170){\vector(0,1){350}}
\put(5,175){\makebox(0,0)[l]{$\remb$}}
\thicklines
\put(-150,150){\line(1,0){300}}
\put(-150,150){\line(1,-1){300}}
\put(150,-150){\line(0,1){300}}
\thinlines
\put(0,150){\line(1,-1){150}}
\put(-50,100){\makebox(0,0){\labelbox{Absolute\\majority\\for $A$}}}
\put(100,100){\makebox(0,0){\labelbox{Absolute\\majority\\for $C$}}}
\put(100,-50){\makebox(0,0){\labelbox{Absolute\\majority\\for $B$}}}
\put(-5,-5){\makebox(0,0)[tr]{0}}
\put(148,-2){\makebox(0,0)[tr]{$1/2$}}
\put(-2,148){\makebox(0,0)[tr]{$1/2$}}
\multiput(-5,-150)(10,0){16}{\line(1,0){5}}
\put(-7,-150){\makebox(0,0)[r]{$-1/2$}}
\multiput(-150,-5)(0,10){16}{\line(0,1){5}}
\put(-150,-7){\makebox(0,0)[t]{$-1/2$}}
\put(-150,150){\circle*{5}}
\put(-150,155){\makebox(0,0)[b]{$A$ unanimous}}
\put(150,150){\circle*{5}}
\put(150,155){\makebox(0,0)[b]{$C$ unanimous}}
\put(150,-150){\circle*{5}}
\put(150,-155){\makebox(0,0)[t]{$B$ unanimous}}
\multiput(0,0)(20,20){8}{\line(1,1){10}}
\put(10,20){\rotatebox{45}{$A$, $B$ tie}}
\multiput(-150,150)(20,-10){15}{\line(2,-1){10}}
\put(5,77){\rotatebox{-26.57}{$B$, $C$ tie}}
\multiput(0,150)(10,-20){15}{\line(1,-2){5}}
\put(58,40){\rotatebox{-63.43}{$A$, $C$ tie}}
\put(73,-2){\makebox(0,0)[tr]{$1/4$}}
\put(-2,73){\makebox(0,0)[tr]{$1/4$}}
\put(-90,75){\rotatebox{-45}{No votes for $C$}}
\put(-55,155){\makebox(0,0)[b]{No votes for $B$}}
\put(155,100){\rotatebox{-90}{No votes for $A$}}
\end{picture}
\end{centre}
\caption{A quasi-impartial culture model for three-candidate elections.}
\label{fig:3emodel}
\end{figure}
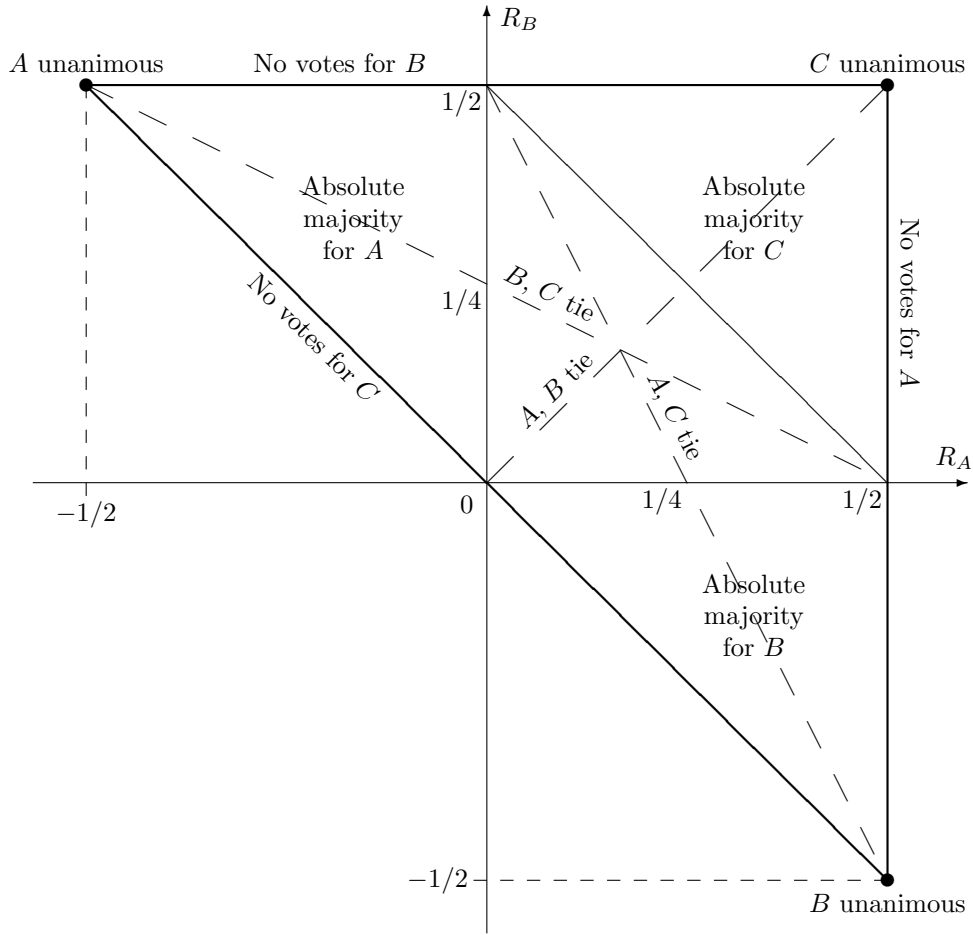

Such a plausible model is illustrated diagramatically in Figure \ref{fig:3emodel}, again based on the remainders $(\rema, \remb)$ for consistency with Figure \ref{fig:winloseset}, although it could just as well be based on the primary vote shares $V_A = a + \alpha$ and $V_B = b + \beta$ which are simple translates of these.
(The equivalent diagram based on $V_A$ and $V_B$ is obtained from Figure \ref{fig:3emodel} by flipping along the diagonal connecting $(1/2,0)$ and $(0,1/2)$, such that the upper-right corner and the origin swap places.)

To explain Figure \ref{fig:3emodel} further: the feasible values of $\rema$ and $\remb$ are individually restricted to the range $[-1/2,1/2]$, where $-1/2$ corresponds to a primary vote of 100\% (unanimity) and $1/2$ to a primary vote of zero.
Hence the universe of feasible pairs $(\rema, \remb)$ is a subset of the square of unit length centred on the origin.
But it must also be the case that $\rema + \remb \in [0,1]$, since this is the primary vote for $C$.
Hence the square is truncated by the main diagonal $\rema + \remb = 0$ from top left to bottom right.
This leaves the right triangle shown with extreme points (vertices) at $(-1/2,1/2)$, $(1/2,-1/2)$ and $(1/2,1/2)$ corresponding to unianimous votes for $A$, $B$ and $C$ respectively and zero votes for the remaining candidates.
Within this larger right triangle lie four smaller congruent right triangles representing the following conditions:
\begin{itemise}
\item
Top left: $\rema < 0$, $\remb > 0$ and $0 < \rema + \remb < 1/2$: candidate $A$ has an absolute majority (and $B$, $C$ have vote shares between 0 and 50\%).
\item
Bottom right: $\remb < 0$, $\rema > 0$ and $0 < \rema + \remb < 1/2$: candidate $B$ has an absolute majority (and $A$, $C$ have vote shares between 0 and 50\%).
\item
Top right: $\rema + \remb > 1/2$, $\rema > 0$ and $\remb > 0$: candidate $C$ has an absolute majority (and $A$, $B$ have vote shares between 0 and 50\%).
\item
Centre: $0 < \rema + \remb < 1/2$, $\rema > 0$ and $\remb > 0$: no candidate has a majority, and all three have vote shares between 0 and 50\%.
(This will be denoted the \emph{region of contest} below.)
\end{itemise}
The \emph{quasi-impartial culture} model is the one where primary-vote probabilities are given by the area measure in Figure \ref{fig:3emodel}, and where voters' second preferences split between the two remaining candidates according to a uniform distribution within a given share of primary votes.
Thus, primary vote scenarios are ascribed equal likelihood if they correspond to equal areas on this diagram.
In particular, each of the three candidates has the same one-in-four chance of winning by absolute majority, and there is a one-in-four likelihood that no candidate attains a majority.

Note this model differs from the usual impartial-culture model that gives equal probability to each voting profile in the hands of an individual voter, in the sense of Section \ref{sec:profiles}.
Under that model, the primary vote $N_A$ for $A$ (for example) will behave as a binomial random variable on $[0,N]$ with parameter $p = 1/3$, which for sufficiently large $N$ can be approximated by a Gaussian variable with mean $N/3$ and variance $2N/9$.
As is well known, the probability that $V_A = N_A / N > 1/2$ giving $A$ a majority will then tend to zero as $N$ increases, at a rate governed by $1/\sqrt{N}$.
(Thus, the likelihood roughly halves for every fourfold increase in the voting population.)

With `impartial culture' in this sense, primary votes will always tend to cluster around the point $V_A = V_B = V_C = 1/3$ (that is, $\rema = \remb = 1/6$ on Figure \ref{fig:3emodel}).
This property is shared by a range of `spatial' models proposed in the social-choice literature (see for example \cite{on:fmfuirv}).
By contrast, the quasi-impartial model drawn from Figure \ref{fig:3emodel} ensures substantial (but not overwhelming) probabilities for majority wins irrespective of the number of voters, which is arguably more typical of real elections.
It goes without saying that no model of this kind should be considered to represent voter behaviour in real elections; they are merely idealised abstractions for study purposes.
But given it is not uncommon in real elections with three contesting candidates for one candidate to receive a majority of primary votes and be immediately elected, it makes sense to have a study model that reflects this.

One may now inquire into the relative likelihood of `winner becomes loser' non-monotonicity under the model just described.
The triangle of potential non-monotonicities in Figure \ref{fig:winloseset} can be readily translated to Figure \ref{fig:3emodel} as a subset of the central `region of contest': two of its sides correspond to parts of the dotted lines labelled `$B$, $C$ tie' and `$A$, $C$ tie' (and represent the loci of points with $\remc = \remb$ and $\remc = \rema$ respectively, where $\remc$ is given by (\ref{eq:remshares})), while the third side is the line segment joining $(1/4,0)$ and $(0,1/4)$.

The area of this triangle is $1/96$, as found above.
But to what `background set' should the area of this triangle be compared?
Proposition \ref{prop:dompair} ensures that in any AV election where no ties occur, there is a unique winner and unique runner-up forming a dominant pair.
Proposition \ref{prop:winlose} deals specifically with elections having $A$ as the winner and $B$ the runner-up, which is only a subset of all possible election scenarios depicted in Figure \ref{fig:3emodel}.
By analysing the possible voter profiles as in the previous section, this subset can be determined as the following:
\begin{proposition}
\label{prop:abprofiles}
An AV election with three candidates $A$, $B$, $C$ has winner $A$ and runner-up $B$ under the following necessary and sufficient conditions.
\begin{enumerate}
\item
There exist numbers $\rema \in [-1/2, 1/4)$ and $\remb \in [0, 1/2)$ with $\rema + \remb > 0$, $\rema + 2\remb < 1/2$ and $2\rema + \remb < 1/2$ such that the primary vote shares for $(A, B, C)$ are $1/2 - \rema$, $1/2 - \remb$ and $\rema + \remb$, respectively (and neither $B$ nor $C$ has as many as half the votes).
\item
The share $\gamma$ of all voters selecting the profile $C > B > A$ is less than $\remb$.
\end{enumerate}
\end{proposition}
The conditions under Proposition \ref{prop:abprofiles} are a strict subset of those under Proposition \ref{prop:winlose}, and stem from a very similar analysis, an important difference being that $\rema$ is no longer required to be positive.
Condition 1, which is equivalent to (\ref{eq:winlose1}), is necessary to ensure that $C$, rather than $A$ or $B$, is the candidate eliminated based on primary votes.
Condition 2, equivalent to (\ref{eq:winlose2}), is necessary to ensure $C$'s preferences elect $A$ rather than $B$.

It is now not difficult to see that the space of primary votes satisfying condition 1 of Proposition \ref{prop:abprofiles} is the kite-shaped region $\rega$ in Figure \ref{fig:3emodel} bounded by the line $\rema + \remb = 0$, the positive $\rema$ axis, and the two `$A$, $C$ tie' and `$B$, $C$ tie' lines.
Outside this region it will be the case either that $B$ attains a majority, or that $C$ has more primary votes than at least one of $A$ or $B$.
By splitting $\rega$ into two triangles along the `$A$, $B$ tie' line it is easy to compute the area of this region as
\begin{equation}
\text{Area of } \rega
   = \half \cdot \frac{\sqrt{2}}{2} \cdot \frac{\sqrt{2}}{6} + \half \cdot \frac{1}{4} \cdot \frac{1}{6}
   = \frac{5}{48}.
\label{eq:abprofarea}
\end{equation}
Since the area of the `universal set'---the original triangle representing all feasible values $(\rema, \remb)$---is $1/2$, the area (\ref{eq:abprofarea}) represents a fraction of $5/24 = 20.83$\% of all feasible values for primary votes.

Note however that the values in this region are only those \emph{necessary} to have $A$ as winner and $B$ as runner-up; they are not \emph{sufficient} without the additional condition 2 on $C$'s preferences.
This will be evident on closer inspection given the region in question includes points for which $\rema > \remb$, and hence for which $B$ has more primary votes than $A$ and will win without an adverse flow of preferences from $C$.
In order to account correctly for the likelihood of satisfying both conditions of Proposition \ref{prop:abprofiles} it is necessary to evaluate a \emph{weighted} area, given by an integral:
\begin{equation}
\text{Weighted area of } \rega = \iint_{\rega} w\paren{\rema,\remb} d\rema d\remb.
\label{eq:abwarea}
\end{equation}
The weight function $w(\rema,\remb)$, for a given primary vote profile, is the relative likelihood of $C$'s preferences satisfying condition 2 of Proposition \ref{prop:abprofiles}.
Since the primary vote share for $C$ is equal to $c + \gamma = 1/2 - \remc = \rema + \remb$, and the quasi-impartial model assumes $\gamma$ has a uniform distribution over its range $[0, \rema + \remb]$, the likelihood that $\gamma < \remb$ given $\rema$ and $\remb$ is
\begin{equation}
w\paren{\rema,\remb} = \frac{\remb}{\rema + \remb} \quad \text{if } \rema > 0, \text{ else 1}.
\label{eq:abweight}
\end{equation}
The weighting is necessarily 1 when $\rema < 0$ since if $A$ has a majority of primary votes the preferences of $C$ cannot elect $B$, and indeed condition 2 is automatically satisfied since $\rema + \remb < \remb$.
Thus the area $\rarea$ being sought is
\begin{equation}
\rarea = \text{Weighted area of } \rega = \rarean + \rareap
\label{eq:abarea}
\end{equation}
where $\rarean$ is the part of $\rega$ with $\rema < 0$ and has unit weighting throughout, while $\rareap$ is the remainder with $\rema > 0$ and has the weighting (\ref{eq:abweight}) at any point $(\rema,\remb)$.
One readily calculates that
\begin{equation}
\rarean = \half \cdot \frac{1}{4} \cdot \half = \frac{1}{16},
\label{eq:abarean}
\end{equation}
while $\rareap$ is calculated in Appendix \ref{sec:rareap} and found to be equal to
\begin{equation}
\rareap = \frac{1}{48}.
\label{eq:abareap}
\end{equation}
Hence
\begin{equation}
\rarea = \frac{1}{16} + \frac{1}{48} = \frac{1}{12}.
\label{eq:abareanp}
\end{equation}
Finally, dividing this by the area $1/2$ of the `universal set' one finds that by elaborate yet rigorous means, one has after all arrived at a simple result.
\begin{proposition}
\label{prop:abfreq}
On the quasi-impartial culture model for three-candidate elections as depicted in Figure \ref{fig:3emodel}, the probability that in an AV election one distinguished candidate $A$ is the winner and another distinguished candidate $B$ is the runner-up is $1/6$.
\end{proposition}
This result is an important justification for the formulation of this model, as it correctly assigns the probability for a distinguished winner and runner-up on an impartial basis.
Obsreve that there are indeed six mutually exclusive ways to select both a winner and runner-up, and on an impartial model these should have equal probability and these probabilities should sum to 1.

One can turn now to the probability of a `winner becomes loser' non-monotonicity on this same model, which will be a similar weighted-area calculation on the triangular set $\triwbl$ depicted in Figure \ref{fig:winloseset}.
The weighting must now consider both conditions 2 and 3 on the second preferences of candidates $B$ and $C$ in Propostion \ref{prop:winlose}.
Once again the share $\gamma$ of $C$ votes with second preference to $B$ can be considered to have a uniform distribution on $[0, \rema + \remb]$, while the share $\beta$ of $B$ votes with second preference to $A$ will have a uniform distribution on $[0, 1/2 - \remb]$.
Thus, the weight to place on any given primary vote profile $(\rema, \remb)$ to ensure conditions 2 and 3 are both met is
\begin{equation}
\weight{\rema}{\remb} = \frac{\remb}{\rema + \remb} \cdot \frac{\rema}{1/2 - \remb}.
\label{eq:wblweight}
\end{equation}
(Observe that conditions such as $\rema < 0$ or $\remb \geq 1/2$ are excluded here by the constraints on $\triwbl$.)
The calculation of the weighted area
\begin{equation}
\twarea = \iint_{\triwbl} \weight{\rema}{\remb} d\rema d\remb
\label{eq:twareaw}
\end{equation}
with $\weight{\rema}{\remb}$ given by (\ref{eq:wblweight}), is carried out in full in Appendix \ref{sec:twarea}.
The result is
\begin{equation}
\begin{split}
\twarea &= \frac{5 \log 2}{8} - \frac{\log 3}{4} + \frac{\log 2 \log 3}{4} - \frac{3 (\log 2)^2}{8}
      - \frac{(\log 3)^2}{8} - \frac{25}{192} + \frac{\pi^2}{48} - \frac{1}{4} \plog\paren{\frac{1}{3}} \\
   &\approx 0.001755535495.
\end{split}
\label{eq:twarea}
\end{equation}
(See Appendix \ref{sec:twarea} for further details.)
This expresses the measure of the set of non-monotonicity scenarios based on Proposition \ref{prop:winlose}, which takes as its prior assumption that the winner $A$ and runner-up $B$ have been identified.
From (\ref{eq:abareanp}) above, the set of all scenarios having winner $A$ and runner-up $B$ has measure $1/12$, and dividing one by the other provides the overall likelihood as now stated.
\begin{proposition}[Likelihood of winner-becomes-loser]
\label{prop:winloseprob}
On the quasi-impartial culture model for three-candidate elections, the likelihood that an election outcome by AV is vulnerable to non-monotonicity of the `winner becomes loser' type is
\begin{equation}
\begin{split}
\frac{\twarea}{\rarea} &= \frac{\log 2}{2} \paren{15 - 9\log 2} - 3 \log 3 \paren{1 - \log 2 + \frac{\log 3}{2}}
      + \frac{\pi^2}{4} - \frac{25}{16} - 3 \plog\paren{\frac{1}{3}} \\
   &\approx 0.021066425942.
\end{split}
\label{eq:winloseprob}
\end{equation}
\end{proposition}
Accordingly, this type of non-monotonicity is latent in just over 2 per cent of election outcomes generated by this model---noting this model fails to give an absolute majority to any candidate in a full one-quarter of cases, which is not necessarily representative of AV elections in practice.

The author has confirmed the result of Proposition \ref{prop:winloseprob} by means of a Monte Carlo simulation.
This samples points $(\rema, \remb)$ uniformly from the triangular `universal set' in Figure \ref{fig:3emodel} to assign primary votes for the three candidates, then samples uniformly within each primary vote share to determine the split of second preferences.
Disregarding ties, the three candidates are then uniquely identified as `winner', `runner-up' or `eliminated' and Proposition \ref{prop:winlose} is used to test vulnerability to non-monotonicity with `winner' in place of $A$, `runner-up' in place of $B$ and `eliminated' in place of $C$.
With sufficient trials (in the order of $10^6$) the proportion of vulnerable outcomes is seen to fall close to the value (\ref{eq:winloseprob}), that is between 2 and $2.2$ per cent.

The result in Proposition \ref{prop:winloseprob} may be compared with that of Lepelley et al \cite{lcb:lompire} who found that the likelihood of a similar vulnerability in three-candidate AV elections was $13/288 \approx 0.04514$, or close to $4.5$ per cent, assuming an `impartial anonymous culture' model as defined in Section \ref{sec:profiles}.
Since this latter model gives equal probability to each equivalent voting profile type as defined in Section \ref{sec:profiles} it also leads to a clustering of election outcomes around the three-way tie $\rema = \remb = \remc = 1/6$, although this clustering will be less strong than under the `impartial culture' model.
The quasi-impartial model as presented here is considered to be more realistic in weighting election outcomes more toward majorities for particular candidates, while still giving substantial weight to outcomes without a primary-vote majority.

\subsection{Loser becomes winner}

Turn now to the `loser becomes winner' variety of non-monotonicity.
In a three-candidate AV election, reasoning as above confirms that a `loser' can only win through a non-monotonic vote shift if they start as the runner-up as part of a dominant pair; after all, if a candidate is eliminated before the final runoff, an adverse shift in preferences against that candidate cannot change that outcome.

Consider then a base scenario as before, with winner $A$, dominant pair $(A,B)$ and a third candidate $C$ who is eliminated before the final runoff, and consider the necessary and sufficient conditions for $B$ to emerge as a winner when voters shift first or later preferences away from $B$.
Since $B$ always loses a runoff against $A$ by assumption (and continues to do so if preferences shift away from $B$), this will again require a shift in the dominant pair, this time from $(A,B)$ to $(B,C)$ with $A$ being eliminated after the vote shift.
This only comes about if primary votes shift to $C$, so that $C$ beats $A$ on primary votes (while ensuring that $B$ beats $A$ also).
Shifting primary votes from $A$ to $C$ so that $A$ is eliminated and loses involves no non-monotonicity, so it is specifically primary vote shfits from $B$ to $C$ that should be studied.

Accordingly, for this analysis let $b' \leq b$ and $\beta' \leq \beta$ denote the share of votes that swing from $B$ first to $C$ first, without regard to how the later preferences shift in these altered ballots.
The rest of the ballots remain unchanged.
As with the `winner becomes loser' scenario, there result six necessary and sufficient conditions on the ballot shares, the first three of which are the same as for that previous analysis since they define the same base scenario:
\begin{itemise}
\item
$a + \alpha < 1/2$, $b + \beta < 1/2$ and $c + \gamma < 1/2$ (no absolute majority in base scenario).
\item
$c + \gamma < \min\paren{a + \alpha, b + \beta}$ ($C$ is eliminated in base scenario).
\item
$a + \alpha + c > 1/2$ and $b + \beta + \gamma < 1/2$ ($A \pref B$ in base scenario).
\item
$(b - b') + (\beta - \beta') > a + \alpha$ ($B$ beats $A$ on primary votes after swing).
\item
$c + \gamma + b' + \beta' > a + \alpha$ ($C$ beats $A$ on primary votes after swing).
\item
$(b - b') + (\beta - \beta') + a > 1/2$ and $c + \gamma + b' + \beta' + \alpha < 1/2$ ($B \pref C$ after swing).
\end{itemise}
On closer inspection one sees that the quantities $b$ and $\beta$ in these conditions always appear as the sum $(b + \beta)$, and similarly for $b'$ and $\beta'$.
In other words, the conditions are agnostic as to which of $B$'s voters shift their votes or what their second preferences are either before or afterwards.
Once again then, one may work with the `remainders' $\rema$, $\remb$ and $\remc$ defined by (\ref{eq:remabc})---which again must all be positive to ensure no majority---and the swing may be encapsulated by the total vote shift
\begin{equation}
\bshift = b' + \beta', \qquad 0 \leq \bshift \leq b + \beta = \half - \remb.
\label{eq:bshift}
\end{equation}
With these definitions the conditions for `loser becomes winner' become
\begin{gather}
\remc > \max\paren{\rema, \remb}, \nonumber \\
c > \rema \qquad \text{and} \qquad \gamma < \remb, \nonumber \\
\rema > \max\paren{\remb + \bshift, \remc - \bshift}, \label{eq:losewin3} \\
a > \remb + \bshift \qquad \text{and} \qquad \alpha < \remc - \bshift. \label{eq:losewin4}
\end{gather}
Again, the first two of these conditions are identical to those for `winner becomes loser', (\ref{eq:winlose1}) and (\ref{eq:winlose2}) respectively.
Combined with the identity (\ref{eq:remshares}) and the positivity constraints, these were seen to imply that $\rema + 2\remb < 1/2$ and $2\rema + \remb < 1/2$, or in terms of Figure \ref{fig:3emodel}, the point $(\rema, \remb)$ lies within the central region bounded by the $\rema$ and $\remb$ axes and the `$B$, $C$ tie' and `$A$, $C$ tie' dotted lines.
In particular $\rema$ and $\remb$ are both individually restricted to the interval $(0, 1/4)$.
Also as before, the condition $\gamma < \remb$ automatically ensures $c > \rema$ since $\rema + \remb = c + \gamma$.

Meanwhile, the two conditions (\ref{eq:losewin4}), that ensure $B \pref C$ after shifting the $\bshift$ votes, also turn out to be equivalent, as a consequence of the identity $a + \alpha = \remb + \remc$.
They also imply the (equivalent) necessary conditions $a > \remb$ and $\alpha < \remc$ prior to giving any (necessarily positive) value to $\bshift$.

Lastly, the condition (\ref{eq:losewin3}) implies that $\rema > \remb$, and that $\bshift$ must satisfy
\begin{equation}
\remc - \rema < \bshift < \rema - \remb.
\label{eq:losewin3a}
\end{equation}
Further, in order for this region for $\bshift$ to be nonempty, it must be the case that
\begin{equation}
2 \rema > \remb + \remc = \half - \rema, \qquad \text{that is,} \qquad \rema > \frac{1}{6}.
\label{eq:losewin3b}
\end{equation}
Finally, with the lower bound $\remc - \rema$ for $\bshift$ the widest possible interval for $\alpha$ by (\ref{eq:losewin4}) is
\begin{equation}
\alpha \in \left[0, \rema\right),
\label{eq:losewina}
\end{equation}
an interval that with $1/6 < \rema < 1/4$ is always nonempty and a proper subset of the interval $[0, 1/2 - \rema]$ where both $a$ and $\alpha$ must actually lie.
Observe that (\ref{eq:losewina}) also implies that $c + \gamma + \alpha = \rema + \remb + \alpha < 2 \rema + \remb < 1/2$, which implies $B \pref C$ \emph{before} the shift as well.

The above yields the following proposition:
\begin{proposition}[Loser becomes winner in AV]
\label{prop:losewin}
Given an AV election having three candidates $(A,B,C)$, a voter profile with $A$ the winner and $B$ the runner-up is vulnerable to non-monotonicity of the `loser becomes winner' type under the following necessary and sufficient conditions.
\begin{enumerate}
\item
There exist numbers $\rema \in (1/6, 1/4)$ and $\remb \in (0, 1/2 - 2\rema) \subseteq (0, 1/6)$ such that the primary vote shares for $(A,B,C)$ are $1/2 - \rema$, $1/2 - \remb$ and $\rema + \remb$, respectively.
In particular, no candidate has an absolute majority, and $A$ must win from a primary vote share greater than $1/4$, less than $1/3$, and less than the primary vote share for $B$ (which is greater than $1/3$).
\item
The share $\alpha$ of all voters selecting the profile $A > C > B$ is less than $\rema$.
\item
The share $\gamma$ of all voters selecting the profile $C > B > A$ is less than $\remb$.
\end{enumerate}
Then for any choice of overall ballot share
\begin{equation}
\bshift \in \paren{\half - 2\rema - \remb, \rema - \remb},
\label{eq:losewin}
\end{equation}
the notional transfer of a $\bshift$ share of ballots with first preference to $B$ to give first preference to $C$ instead (regardless of second preferences before or after) will cause $B$ instead of $A$ to be elected.

The above conditions ensure that the interval in (\ref{eq:losewin}) is a nonempty subset of $[0, 1/2 - \remb]$.
They also ensure that $A \pref B$ and $B \pref C$ throughout, so that either a Condorcet cycle exists or $A$ is the Condorcet winner.
\end{proposition}
Observe that by choosing $\rema$ close to $1/4$, or $\remb$ close to the upper limit $1/2 - 2\rema$ in condition 1, candidates $A$ and $C$ are brought very near to a tie for second place behind $B$, with $A$ remaining just in front ensuring $C$ is eliminated.
The shift of votes $\bshift$ required to promote $B$ to a winning position instead of $A$ can then be brought arbitrarily close to zero, given a large enough voting population.

The mechanism for non-monotonicity here is seen to be that $A$ is strongly favoured to win by $C$'s voters despite falling below $B$ on primary votes, so that $A$ wins as long as their primary vote remains ahead of $C$'s, otherwise $B$ wins on $A$'s preferences.
In practical terms, it is as though $A$ and $C$ form a coalition against $B$, but with more `preference leakage' to $B$ from $A$ than from $C$, making it strategic for $B$ voters to boost $C$'s vote and ensure $A$'s elimination.

\begin{figure}[t]
\begin{centre}
\begin{picture}(350,250)(-25,-25)
\put(-5,-5){\makebox(0,0)[tr]{0}}
\put(-25,0){\vector(1,0){350}}
\put(325,-10){\makebox(0,0)[t]{$\rema$}}
\put(200,-10){\line(0,1){10}}
\put(200,-15){\makebox(0,0)[t]{$1/6$}}
\put(300,-10){\line(0,1){20}}
\put(300,-15){\makebox(0,0)[t]{$1/4$}}
\put(0,-25){\vector(0,1){250}}
\put(-10,325){\makebox(0,0)[r]{$\remb$}}
\put(-10,200){\line(1,0){10}}
\put(-15,200){\makebox(0,0)[r]{$1/6$}}
\put(200,0){\line(0,1){200}}
\put(188,100){\rotatebox[origin=c]{-90}{$\rema = 1/6$}}
\put(200,200){\line(1,-2){100}}
\put(230,110){\rotatebox[origin=b]{-63.43}{\labelbox{$2\rema + \remb = 1/2$\\$(\remc - \rema = 0)$}}}
\put(200,200){\circle*{5}}
\put(205,205){\makebox(0,0)[bl]{\labelbox{Three-way tie\\$\rema = \remb = \remc = 1/6$}}}
\multiput(0,200)(20,0){10}{\line(1,0){10}}
\multiput(200,0)(0,20){10}{\line(0,1){10}}
\end{picture}
\end{centre}
\caption{Visualisation of profiles $(\rema,\remb)$ vulnerable to non-monotonicity under Proposition \ref{prop:losewin}.}
\label{fig:losewinset}
\end{figure}
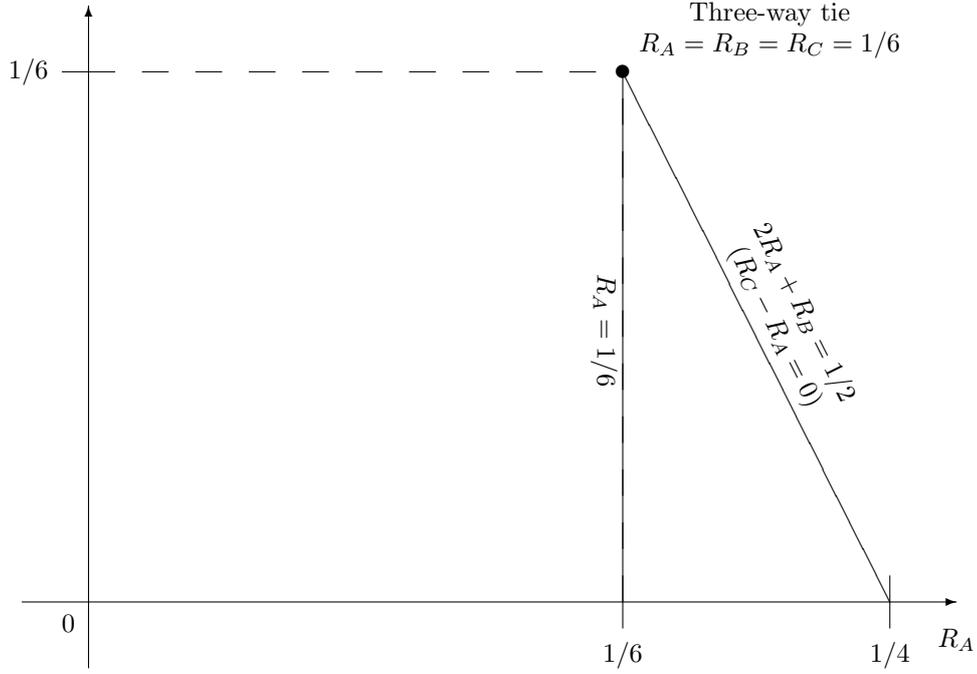

Figure \ref{fig:losewinset} is analogous to Figure \ref{fig:winloseset}, and illustrates the set of profiles that are potentially vulnerable to `loser becomes winner' under Proposition \ref{prop:losewin} based on the three candidates' primary votes alone.
This point set is the interior of the right triangle defined by the inequalities $\rema > 1/6$, $\remb > 0$ and $\remb < 1/2 - 2\rema$ (equivalently, $\remc > \rema$ with $\remc$ implicitly defined by (\ref{eq:remshares})), and satisfying condition 1 of Proposition \ref{prop:losewin}.
Voting profiles within this set must also satisfy the two conditions on the second preferences of $A$ and $C$ voters.

Denoting this right triangle $\trilbw$, this is readily seen once again to be a subset of the region $\rega$ where it is possible to have candidate $A$ as winner and $B$ as runner-up.
It also partially overlaps the set $\triwbl$ depicted in Figure \ref{fig:winloseset} where profiles are potentially vulnerable to `winner becomes loser'.
Its `gross area' is readily calculated as $1/144$, making it smaller than $\triwbl$ in gross terms.
Thus, considering primary votes alone, outcomes potentially vulnerable to `loser becomes winner' are less likely to occur than those vulnerable to `winner becomes loser'.

Once again, one may determine the likelihood of outcomes with `loser becomes winner' vulnerabilities by calculating the weighted area
\begin{equation}
\tlarea = \iint_{\trilbw} \weight{\rema}{\remb} d\rema d\remb
\label{eq:tlareaw}
\end{equation}
where now the weighting required to assign uniform probability to second preferences satisfying Proposition \ref{prop:losewin} is
\begin{equation}
\weight{\rema}{\remb} = \frac{\remb}{\rema + \remb} \cdot \frac{\rema}{1/2 - \rema}.
\label{eq:lbwweight}
\end{equation}
Observe that condition 3 for Proposition \ref{prop:losewin} is the same as that for Proposition \ref{prop:winlose}, so generates the same weighting factor, while condition 2 calls for a quantity sampled from a uniform distribution on $[0, 1/2 - \rema]$ to be less than $\rema$, leading to the second factor in (\ref{eq:lbwweight}).

The calculation of the weighted area $\tlarea$ by (\ref{eq:tlareaw}) is carried out in Appendix \ref{sec:tlarea}, with the result
\begin{equation}
\begin{split}
\tlarea &= \frac{1}{18} + \frac{11}{72} \log 2 - \frac{\log 3}{8} - \frac{\paren{\log 2}^2}{8}
      + \frac{\paren{\log 3}^2}{8} - \frac{\pi^2}{48} + \frac{1}{4} \plog\paren{\frac{1}{3}} \\
   &\approx 0.000875047890957,
\end{split}
\label{eq:tlarea}
\end{equation}
and dividing this by the area $1/12$ of the set $\rarea$ of scenarios having $A$ as winner and $B$ as runner-up gives the following:
\begin{proposition}[Likelihood of loser-becomes-winner]
\label{prop:losewinprob}
On the quasi-impartial culture model for three-candidate elections, the likelihood that an election outcome by AV is vulnerable to non-monotonicity of the `loser becomes winner' type is
\begin{equation}
\begin{split}
\frac{\tlarea}{\rarea}
   &= \frac{2}{3} + \frac{11}{6} \log 2 + \frac{3}{2} \brak{\log 3 \paren{\log 3 - 1} - \paren{\log 2}^2}
      - \frac{\pi^2}{4} + 3 \plog\paren{\frac{1}{3}} \\
   &\approx 0.0105005746915.
\end{split}
\label{eq:losewinprob}
\end{equation}
\end{proposition}
Thus overall, just over 1\% of outcomes in the quasi-impartial culture model suggested by Figure \ref{fig:3emodel} are vulnerable to `loser becomes winner'---a little less than half as many as are vulnerable to `winner becomes loser'.

Once again, this likelihood figure can be verified through a Monte Carlo simulation, sampling at random from the space of primary votes $(\rema, \remb)$ of Figure \ref{fig:3emodel}, identifying the winner and runner-up, then assigning second preferences for each group of voters by sampling appropriate uniform distributions.
When this is done, the relative frequency with which the sampled outcomes satisfy the conditions of Proposition \ref{prop:losewin} (suitably interpreted in terms of the winner, runner-up and eliminated candidate in each case) is found to fall between 1\% and 1.1\%, consistent with Proposition \ref{prop:losewinprob}.

\subsection{Simultaneous non-monotonicity: a rare curiosity}

A simple comparison of Figures \ref{fig:winloseset} and \ref{fig:losewinset} reveals that the two regions of vulnerable primary vote profiles have a nonempty intersection.
It is therefore possible that within this intersection there are voting profiles that satisfy the conditions of Proposition \ref{prop:winlose} and Proposition \ref{prop:losewin} simultaneously.
In such situations---again with $A$ the winner and $B$ the runner-up without loss of generality---it is possible to shift votes from $B$ to $A$ such that $A$ loses the election to $C$, and also from the same starting point, to shift votes from $B$ to $C$ such that $A$ loses the election to $B$.
Thus, the same scenario is vulnerable to \emph{both} varieties of non-monotonicity simultaneously.
(See also \cite{ft:idmfwdoiufvm}.)

In principle, one can ascertain the likelihood of this simultaneous non-monotonicity in the quasi-impartial model by calculating the weighted area of the intersection
\begin{equation}
\tsarea = \iint_{\triwbl \cap \trilbw} \weight{\rema}{\remb} d\rema d\remb
\label{eq:tsareaw}
\end{equation}
with a weighting based on the likelihood of satisfying all three additional conditions (conditions 2 and 3 of Proposition \ref{prop:winlose} and condition 2 of Proposition \ref{prop:losewin}):
\begin{equation}
\weight{\rema}{\remb}
   = \frac{\remb}{\rema + \remb} \cdot \frac{\rema}{1/2 - \remb} \cdot \frac{\rema}{1/2 - \rema}.
\label{eq:simweight}
\end{equation}
The calculation of $\tsarea$ is left as an exercise for the reader.
Empirically, Monte Carlo simulation reveals that the fraction $\tsarea / \rarea$ of election outcomes in the model that satisfy all conditions simultaneously is likely between $0.3$\% and $0.4$\%.
Thus it is conjectured that $0.003 < \tsarea / \rarea < 0.004$.

Note that a consequence of having scenarios vulnerable to both non-monotonicities simultaneously is that the likelihood of being vulnerable to one or other non-monotonicity is strictly less than that given by the sum $\twarea + \tlarea$.
Instead, it is given by the inclusion-exclusion principle as
\begin{equation}
\text{Likelihood of either non-monotonicity} = \frac{\twarea + \tlarea - \tsarea}{\rarea}.
\label{eq:nonmonu}
\end{equation}
Assuming $\tsarea / \rarea$ is $0.3$\% or greater, this joint likelihood will be somewhat less than 3\% of all three-candidate election outcomes generated by the model.

A concrete demonstration is warranted at this point.
By inspection of Figure \ref{fig:winloseset} and Figure \ref{fig:losewinset}, it is apparent that the point $(\rema, \remb) = (9/48, 1/12)$ represents a `typical' primary vote scenario satisfying the first conditions for both winner-becomes-loser and loser-becomes-winner.
The value $\remb = 1/12$ mediates between the minimum and maximum allowable values for $\remb$ from Figure \ref{fig:losewinset}, while $\rema = 9/48$ falls in the middle of the cross-section $\rema \in (1/6, 5/24)$ formed by the intersection of the line $\remb = 1/12$ with either set in Figure \ref{fig:winloseset} or Figure \ref{fig:losewinset}.
The corresponding primary vote shares are $1/2 - 9/48 = 15/48$ for $A$, $1/2 - 1/12 = 5/12$ for $B$ and $9/48 + 1/12 = 13/48$ for $C$.

Regarding the distribution of second preferences, observe that the conditions for both Proposition \ref{prop:winlose} and Proposition \ref{prop:losewin} will be met if all $A$ voters give their second preference to $B$, all $B$ voters give theirs to $C$, and all $C$ voters give theirs to $A$.
Table \ref{tab:nonmon} provides a concrete example of this `disciplined preference' scenario under the column `Election 1' with a population of 4800 voters.
\begin{table}
\begin{centre}
\begin{tabular}{cc|r|r}
& & \multicolumn{2}{c}{Votes} \\
Type & Ballot order & Election 1 & Election 2 \\ \hline
$a$ & $A > B > C$ & 1500 & 700 \\
$\alpha$ & $A > C > B$ & 0 & 800 \\
$b$ & $B > C > A$ & 2000 & 1200 \\
$\beta$ & $B > A > C$ & 0 & 800 \\
$c$ & $C > A > B$ & 1300 & 1000 \\
$\gamma$ & $C > B > A$ & 0 & 300 \\ \hline
\multicolumn{2}{r|}{Total ballots} & 4800 & 4800
\end{tabular}
\end{centre}
\caption{Examples of three-candidate AV elections exhibiting dual non-monotonicity.}
\label{tab:nonmon}
\end{table}
The `Election 2' column, meanwhile, provides an example with a more `typical' distribution of preferences, flowing from each candidate to both other candidates with the proportions $\alpha$, $\beta$ and $\gamma$ set to just below their maximum values for non-monotonicity.
In both elections, $A$ is elected on preferences from $C$.

Looking first at Election 1 in Table \ref{tab:nonmon}, it is easy to see that any hypothetical transfers of votes between candidates $A$ and $C$ are monotonic in their effect: either leaving the outcome unchanged, or (when more than 100 votes transfer from $A$ to $C$) causing $A$ to be eliminated and electing either $B$ or $C$.
Likewise, a transfer of votes to $B$ from either $A$ or $C$ is neither detrimental to $B$ nor advantageous to $C$, so does not violate monotonicity either.
It is only transfers of votes from $B$ that \emph{might} be non-monotonic, in either of two directions: if to the benefit of $A$ by between 701 and 899 votes it causes $C$ to win instead of $A$; and to the benefit of $C$ by between 201 and 499 votes it results in $B$ winning instead of $A$.
If the transfer of votes from $B$ falls outside these bounds, the outcome remains monotonic (or results in a tie).

All the same observations can be made regarding Election 2; here it helps to recall Proposition \ref{prop:dompref}, by which the only second preferences that are relevant to the outcome are those for the eliminated candidate.
It does mean, however, that when it comes to shifting votes from $B$ to $A$ to produce a non-monotonic loss for $A$, the number that must be shifted is between 701 and 899 as for Election 1, but now no more than 99 of these can be of ballot type $b$ ($B > C > A$), otherwise there are insufficient second preferences to $C$ from the remaining votes for $B$ to elect $C$ and turn $A$ into a loser after $B$ is eliminated.
Meanwhile, to produce a non-monotonic win for $B$ by shifting votes from $B$ to $C$ it suffices to shift between 201 and 499 votes without regard to preferences since the candidate eliminated after the shift ($A$) is not involved in the transfer of votes.

It may be noticed that the cyclic flow of preferences selected for the `Election 1' example above also brings about a Condorcet cycle with $A \pref B$, $B \pref C$ and $C \pref A$---evidently, the elimination of any one of $A$, $B$ or $C$ sends their entire vote to the next candidate in the cycle.
The same is also true, albeit less obviously, of `Election 2'.
That a Condorcet cycle occurs in both examples is no coincidence, as formally stated in the following proposition which is an immediate corollary of Propositions \ref{prop:winlose} and \ref{prop:losewin}.
\begin{proposition}[Dual non-monotonicity is a Condorcet cycle]
\label{prop:nonmon}
Any AV election with three candidates---without loss of generality identified as winner $A$, runner-up $B$ and eliminated $C$---that simultaneously meets the conditions of Proposition \ref{prop:winlose} and Proposition \ref{prop:losewin}, and therefore exhibits both varieties of non-monotonicity simultaneously, is a Condorcet cycle with $A \pref B$, $B \pref C$ and $C \pref A$.
\end{proposition}

\subsection{Summary: what does non-monotonicity mean?}

Given all the aforementioned findings, what can be said about the effect of non-monotonicity on the ability of the AV election procedure to reflect the will of an electorate in electing a single winner?

The first conclusion that may be drawn is a practical one: instances of non-monotonicity are expected to be rare.
When the casting of ballots by voters is modelled by the quasi-impartial culture framework suggested by Figure \ref{fig:3emodel}, it is found that fewer than 3\% of possible elections are vulnerable to any kind of non-monotonicity.
But even this is practically certain to overestimate the likelihood in real elections, where genuine three-candidate contests are a great deal less common than the model suggests.

At a deeper level though, non-monotonicity in AV is also demonstrative of the problematic nature, from a social-choice perspective, of winner-take-all contests in which there is neither a majority winner nor a clearly defined dominant pair, by virtue of there being three candidates each with more than one-quarter of primary votes.
It is in this space that non-monotonicities reside, and there is also a clear intersection with the presence of Condorcet cycles and related `instabilities'.
Thus it has been found that, for all three-candidate scenarios vulnerable to either non-monotonicity, one of the following is always true:
\begin{itemise}
\item
there is a Condorcet cycle, meaning no candidate can be said to have a majority mandate in purely \emph{relative} terms; or
\item
there is a Condorcet winner who achieves less than one-third of primary votes, and so is elected only conditionally by AV.
(See Proposition \ref{prop:condorcet}.)
\end{itemise}
In the latter case, one also observes that wherever non-monotonicity occurs, it involves a hypothetical exchange of votes between the two candidates who are \emph{not} Condorcet winners.
Further, non-monotonicity can arise in only one of the two possible senses, and is found on further investigation to be one of the following two species:
\begin{itemise}
\item
winner becomes loser: here the `winner' is not the Condorcet winner, and when they become a `loser' it is as a consequence of the Condorcet winner being elected;
\item
loser becomes winner: the `loser' is so by virtue of the Condorcet winner being elected, and becomes a `winner' by beating the Condorcet winner into third place, in the sense discussed in Section \ref{sec:condorcet}.
\end{itemise}
It was also seen above that whenever an election scenario is vulnerable to non-monotonicity, the imposition of a single additional condition on voters' second preferences is sufficient to force a Condorcet cycle, as per Proposition \ref{prop:nonmon}.
Accordingly for such scenarios, even when there is a Condorcet winner there is a sense in which this  winner is `weak' and the scenario `close' to a Condorcet cycle.
The following proposition, applicable more generally and not just in non-monotonic cases, makes this idea more rigorous.
The proof is given in Appendix \ref{sec:condinstab}.
\begin{proposition}[Condorcet instability in AV elections]
\label{prop:condinstab}
If any AV election has a Condorcet winner $C$ who is not elected, then there exists a modification to the voter profile that converts the election to a top Condorcet cycle without changing any primary votes.
Further, this modification can be done in a way that does not alter the election outcome, nor the result at any stage of the AV counting process, and is therefore of necessity monotonic.
If there are just three candidates, or the elimination of Condorcet winner $C$ occurs only at the final stage of the count leaving just two contesting candidates, then there is a modification to a Condorcet cycle that alters only the later preferences of the candidate that is eventually elected.
\end{proposition}
So it transpires that even where a non-monotonicity in three-way AV contests does not herald a Condorcet cycle, it nonetheless signals a scenario that becomes a Condorcet cycle with a (monotonic) change in one candidate's non-exercised second preferences.

This pervasive linkage of non-monotonicity with Condorcet cycles is instructive.
The Condorcet cycle is the \emph{sine qua non} of voting paradoxes for single-winner elections as it is latent in \emph{any} procedure that asks voters to rank their preferences for candidates, and stems from an impasse within the voters' own choices, not a contingent property of the voting procedure.
Any method for extracting a single winner from a Condorcet cycle of voter preferences will be debatable, and arguably this extends to the case of an election that would be a Condorcet cycle but for a fortuitous accident within one candidate's preferences.

Non-monotoniciity is a rare quirk in the AV voting system, occuring in rather special conditions that are unlikely in practice owing to the low incidence of genuine three-cornered contests.
It has been shown that this phenomenon is also intimately linked with the bane of \emph{all} rank-based voting systems, the Condorcet cycle.
It is, ultimately, another phenomenon that highlights the tension between contested ideas of how `majority rule' is best expressed in a winner-takes-all election; ideas that AV ultimately attempts to hold in the balance rather than adopt one notion to the exclusion of others.

\section{STV and proportional representation}
\label{sec:proportional}

Unlike in winner-takes-all contests, in multi-winner elections there is the opportunity to elect winning candidates with sectional constituencies whose diversity reflects the plurality of views within the electorate.
The usual means of reconciling the fundamental notions of `majority rule' and `voter choice' in the multi-winner context is through the principle of \emph{proportional representation}: that the diversity in elected candidates should reflect the diversity of the electorate according to primary vote proportions.

A variety of `party list' systems are in existence and are used in numerous jurisdictions to achieve proportional representation by having voters make (in the simplest cases) a single selection from a set of party lists, with an ingenious arithmetic procedure used to apportion winning candidates from these lists according to the total vote directed toward each.
The operation of such systems can be likened to optimisation procedures, that minimise the `deviation' from ideal proportionality according to a suitably formulated performance index.

The STV procedure, as is well known, also achieves proportional representation among stable coalitions, when voters in effect have a party ranking in mind, and then rank candidates in order within party blocks, the blocks following the overall party ranking.
This party-ranking behaviour in STV elections is encouraged by the current Senate voting system in Australia.
But at the basic level the STV procedure does not actually require candidates to formally coalesce into parties, but allows coalitions to emerge through the will of the voters themselves, thus arguably foregrounding the principle of `voter choice'.

Yet the idea that STV is a system of proportional representation at all is challenged by at least one theorist, who contends that it is so only through a kind of contingent sleight of hand \cite{m:uaes}:
\begin{quotation}
The only electoral procedure that really implements [deterministic proportional representation, PR] (as far as possible) is the list system\ldots
The STV procedure is often claimed by politicians and journalists to be a PR system.
But social-choice theorists know very well that this claim is incorrect.
This is not only easy to prove in theory (for example, by observing that STV is not monotonic), but can also be seen in practice by examining the results of elections conducted under STV.
In fact, STV is a [district representation] system that is ingeniously designed to produce less disproportionate outcomes than the extremely pathological plurality procedure.
However, the approximate degree of proportionality it produces is quite erratic.
In particular, STV is biased against small and radical parties.
\end{quotation}
The writer here appears to have been strongly influenced in their opinion by the outcomes of Irish parliamentary elections, which are conducted by STV, but have historically evolved in such a way that many individual contests have as few as three winners.
As is also quite well known, the fewer the number of winners in a STV election (as indeed with a list system) the more it takes on the characteristics of a winner-takes-all contest and works against candidates with a minority of primary votes.
Generally in Australia where STV elections are conducted, except in some local government elections and Territory Senate elections, the number of winners in one election (commonly referred to as \emph{district magnitude}) is set no lower than five.

The following principle extends Proposition \ref{prop:avcoalition} and formalises the mechanism by which STV achieves proportional representation.
\begin{proposition}[Proportionality for stable coalitions in STV]
\label{prop:stvcoalition}
In a STV election with $W$ winners, $N$ ballots and a quota $Q$ calculated accordingly, if there exists a subset (`coalition') $C$ of candidates such that $V \geq Q$ voters rank all candidates in $C$ ahead of any candidate outside $C$, then:
\begin{enumerate}
\item
if $C$ contains $\floor{V / Q}$ or more candidates, then STV will elect at least $\floor{V / Q}$ from $C$; and
\item
if $C$ contains fewer than $\floor{V / Q}$ candidates, then STV will elect all candidates in $C$.
\end{enumerate}
More than once such subset $C$ may exist in a STV election.
\end{proposition}
This result is robust against the presence of exhausted votes that reduce the notional (if not actual) quota $Q$, and also against the small notional change in $Q$ that follows the partial election of candidates by Proposition \ref{prop:stvquota}.
Such changes are always such as to \emph{reduce} the notional quota, which if fully taken into account (as in some STV variants, through not those in use in Australia) might potentially increase the number of elected candidates from $C$ but never reduce it from the number $\floor{V / Q}$ guaranteed by Proposition \ref{prop:stvcoalition} (again, assuming $C$ contains sufficient candidates).

Proposition \ref{prop:stvcoalition} works because among the $V$ ballots that prefer candidates in $C$, all surplus transfer votes, as well as votes from eliminated candidates, must continue circulating among the candidates in $C$ before passing to any others.
They do so with a combined value equivalent to at least $\floor{V / Q}$ quotas, less $Q$ for each elected candidate, and will not fall below a single quota $Q$ of value before either electing all candidates in $C$ (if $|C|$ is less than $\floor{V / Q}$), or electing $\floor{V / Q}$ candidates in $C$.
Only after falling below a single quota, or exhausting the set $C$, will these ballots transfer outside $C$ due to election or elimination.
Note also that if $\floor{V / Q}$ equals or exceeds the number of candidates in $C$, it will never come about that candidates in $C$ are eliminated: the STV procedure ensures that candidates with a full quota are elected before any elimination takes place, and in this case there will \emph{always} be at least one candidate in $C$ with a full quota until all in $C$ are elected.

(At the same time, Proposition \ref{prop:stvcoalition} does \emph{not} extend to a guarantee that STV will elect $\floor{V / Q}$ candidates from $C$ where $V$ is simply the number of ballots that transfer to candidates in $C$ after elimination of candidates \emph{outside} $C$.
There have been real-world cases where parties have been denied at least one elected position due to their candidates being eliminated before receiving the benefit of preferences from other parties' voters.
This includes cases where it is possible a party could have tactically benefited by running \emph{fewer} candidates, in the hope of garnering higher individual primary votes for the smaller number and influencing the order of elimination in their favour \cite{m:fvipsfw}.
Importantly, these tactical scenarios relate only to parties seeking support \emph{outside} stable coalitions in the sense of Proposition \ref{prop:stvcoalition}, and while this sometimes produces outcomes that are unfortunate for \emph{parties}, it is far less clear that the will of \emph{voters} is in any way thwarted by such results.)

Proposition \ref{prop:stvcoalition} ensures that if all voters cast their ballots party-by-party, in particular putting all candidates from their favoured coalition first (regardless of the ordering of individual candidates), then STV operates in essentially the same manner as a party-list system on the so-called `largest remainder' principle.
This type of party-list system simply divides each party's vote by $Q$ and rounds down to determine the number elected (quotient), before filling remaining positions based on the remainders after division.
In many cases all but perhaps one of the available positions are filled by quotients in this manner; then the two systems will only differ in how the last winner is decided, with the greatest-remainder system essentially conducting a FPTP election with the surplus votes, while STV conducts an AV election for the final position.

The largest-remainder principle is clearly proportional in its outcomes at least in a broad sense, although when it is used in party-list systems it is generally the Hare quota $Q = \floor{N / W}$ that is preferred theoretically, rather than the Droop quota $Q = \floor{N / (W + 1)} + 1$ according to (\ref{eq:droop}).
This is because results using the Hare quota are \emph{unbiased} in the sense that, on average across all feasible party-list election scenarios, the proportion of elected positions allocated to each party is equally likely to be greater than, or less than, the proportion of votes cast for each party \cite{p:pramta}.
When the Droop quota is used in party-list systems, on the other hand, it is found to be more likely that large parties are allocated a slightly larger share of elected positions, and small parties a slightly smaller share, than given ideally by their exact share of the vote.

Largest-remainder list systems using the Hare quota are also found to optimise the \emph{Loosemore--Hanby index}, essentially the total absolute deviation of parties' share of winners from their share of votes \cite{lh:tlomd}:
\begin{equation}
D = \half \sum_{k=1}^p \abs{\frac{E_k}{W} - \frac{V_k}{N}},
\label{eq:lhindex}
\end{equation}
where $p$ is the number of parties, $E_k$ the number of elected positions from party $k$ and $V_k$ the number of ballots cast for party $k$.
Loosemore and Hanby's original paper \cite{lh:tlomd} found that the index (\ref{eq:lhindex}) generally took lower values for lowest-remainder apportionment than for other systems (specifically the d'Hondt and Sainte-Lagu\"{e} systems which employ the `method of highest averages'), without giving any specific quota formula.
(Meanwhile there are alternative indices to (\ref{eq:lhindex}), based on squared rather than absolute differences, and at least one of these is optimised by the Sainte-Lagu\"{e} method.)

In the context of party-list systems, proportionality as measured by (\ref{eq:lhindex}) and similar indices is the clearest expression of the common principle of `one vote one value' given the will of the voter is expressed almost solely through the selection of a single party list.
STV's use of the Droop quota means that in this sense it is not `optimal' according to any of these indices.
On the other hand it enjoys the following features that distinguish it from party-list systems:
\begin{enumerate}
\item
Because voters can freely rank candidates rather than parties, and can choose candidates across multiple parties, coalition formation is a process that emerges from the will of the voters themselves rather than being imposed \emph{a priori} through the formalisation of parties.
Although this is taken as a disadvantage of STV when the focus is on the `correctness' of apportionment of winners by party (see above), it can readily be argued that STV better reflects the principle of `voter choice'.
\item
Voters whose first preference is for a minority party or candidate who is not assured of achieving a quota $Q$ of votes can nominate alternative candidates through their ranked preferences and thereby ensure with high probability that at least one of their nominated candidates is elected.
\item
Since the Droop quota is smaller than the Hare quota, minor parties have a greater chance of achieving a single quota and thereby winning from a small but significant share of votes.
Further, since the final winning position is in effect determined by an AV election, a loose coalition of minor party and independent candidates can recommend mutual ranking to voters and thereby improve their chances of winning through the operation of Proposition \ref{prop:avcoalition}.
\end{enumerate}
The above can be further illustrated through some remarks on Australian Senate elections.
In a typical half-Senate election, each state in Australia elects six senators by STV, and accordingly the usual quota for a Senate seat is 14.3\% of the vote (in round figures).
The pattern of voting in Australian elections in recent times has been for the two major political parties (strictly speaking coalitions) to each glean approximately one-third of primary votes, with the remaining one-third going to `crossbench' candidates (minor parties and independents).
In the case of the Senate, this usually operates to elect two Senators for each state from each of the major parties, with the remaining two Senate seats in each state contested by a wider range of candidates.
This contrasts with the situation for much of the 20th century when minor candidates had little support, and the two major parties gained close to half the vote each and were commonly assured of three Senators each per state.
(A stable coalition wins three of six Senate seats with just 43\% of primary votes.)

Minor parties had emerged to challenge the two-party dominance by 1984, when the `above the line' voting option was introduced, with the aim of simplifying the voting process given the burgeoning number of small parties and candidates running.
This, for practical purposes, converted the process to a closed party-list system---albeit one where votes would follow each party's nominated ranking (or `group ticket') all the way down through the \emph{full} list of candidates.
While the option remained to cast a traditional STV ballot `below the line', this option was exercised by only a small percentage of voters in practice, as a valid `below the line' ballot required an error-free ranking of all, or nearly all, of several dozen individual candidates.

As micro-party Senate candidates continued to grow in number and variety through the 2000s and early 2010s it became increasingly questionable, with political parties rather than voters themselves determining the ranking of these micro-party candidates in the vast majority of ballots, whether the outcome of Senate elections was genuinely reflecting the will of voters rather than skilful horse-trading among party officials.
This was particularly so in the contest for the final place, which was by then almost always contested by minor and micro-party candidates given the major parties' votes by that time fell well short of the 43\% necessary to gain a third quota.
It is likely that several of the Senate crossbenchers elected during this era would never have been elected had they relied on being ranked by voters themselves, rather than by the group tickets of parties that gained more votes.

A 2016 reform accordingly abolished the system of group tickets, in favour of a system where voters themselves rank either parties `above the line' or individual candidates `below the line'.
The requirement to cast a full ranking of candidates `below the line' was also discarded, so that voters are able to stop after a minimum number calibrated to the number of winning positions.
In practice the system thus resembles an open party-list system, but with more flexibility for voters to rank individual candidates from different parties in arbitrary order.
The results of Senate elections since 2016 have resembled less a lottery among little-known micro-parties and more a selection of crossbench candidates that are each found to have small but substantial support in their own right.
There has also arguably been a decline in the number of parties seeking election, with the various micro-parties organising themselves into more explicit coalitions under a single party banner, in order to gain the public profile required to garner genuine support within the electorate at large.

At the time of writing, group ticket voting in Australia survives only in upper-house elections in the state of Victoria, where an active campaign seeks reform of this along the same lines as for the Senate above.
Other Australian states that employed this voting system for their upper-house elections have already followed the Senate example, most recently Western Australia.
Thus STV in the Australian context has been steadily evolving into a form that combines practicality with a more explicit recognition of voter choice, maintaining a suitable degree of proportionality in the allocation of parliamentary seats to the larger parties while ensuring smaller parties and coalitions are represented in accordance with voters' own allocated preferences.

\section{Conclusion}
\label{sec:conclu}

When it comes to voting in elections---particularly in those for which there can only be one winner---people can rationally hold contesting ideas of how self-evident but informal principles such as `majority rule', `voter choice' and `one person, one vote' are best expressed in concrete situations.
On the one hand, there is the clear mandate expressed by a voter giving their `number one' preference to a specific candidate.
On the other hand there is Condorcet's principle by which a candidate most favoured by the electorate should be seen to win as many one-on-one runoffs as possible against the other candidates, irrespective of whether or not voters place them first on their ballots.

There are of course many happy circumstances where the competing desiderata are in full agreement, as for example where one candidate wins an absolute majority of primary votes.
But when subjected to the vast motley kaleidoscope of election scenarios that lie beyond the obvious, not only do the various desiderata differ in their conclusions, but there is no single desideratum that goes entirely unchallenged.
This is notoriously the case with relying on primary-vote mandates alone as in FPTP, rife as this is with spoiler effects, the elevation of large minorities over genuine majorities, and the effective barring of new entrants to the political process that do not have the endorsement of established parties.
But even the venerable Condorcet winner principle can be questioned in certain cases, as for example when it compels the election of a candidate that not one voter has credited with a first-place ranking \cite{rg:pppidav}.
The ever-present potential for Condorcet cycles also frustrates in principle the idea of relying on one-on-one runoffs as the sole criterion for election.

Transferable voting systems---Alternative Vote for winner-takes-all elections and its generalisation to Single Transferable Vote for multi-winner elections---have been defended against various `Condorcet consistent' alternatives put forward by theorists on purely \emph{pragmatic} grounds, noting for example the difficulty of conducting exhaustive two-candidate runoffs to identify a Condorcet winner, and the contrasting simplicity with which a dominant pair can frequently be identified for a quick runoff result in an AV election.
But while such considerations are important in practice, equally important is that there exist reasonable \emph{philosophical} arguments for the use of such systems: not least, that they strike a necessary balance between the competing desiderata for identifying socially preferred candidates.

On the `plurality mandate' side, AV elections ensure that where there are two clearly identifiable `front runner' candidates each with a strong showing in primary votes, the result will be determined by a runoff between them (Proposition \ref{prop:domprim}).
AV will ensure the election not only of an individual candidate who wins an absolute majority of primary votes, but also one from any stable coalition of candidates whom a majority of voters prefer to others (Proposition \ref{prop:avcoalition}); in this way AV is protective against attempts to influence elections by running additional `clone' candidates.
Meanwhile, on the `Condorcet comparison' side, AV ensures that the result is \emph{always} that of a two-candidate runoff, beween the ultimate winner and a credible runner-up (Proposition \ref{prop:dompair}).
For this reason, AV never elects a Condorcet loser.
Further it guarantees the election of a Condorcet winner provided they garner at least one-third of primary votes or preferences while still contesting (Proposition \ref{prop:condorcet}).

Nonetheless, AV like all voting systems has its troublesome cases.
The two greatest flaws generally held against AV are on one hand, its failure to elect a Condorcet winner when two other candidates have strong support nearer the top of voters' ballots, and on the other, the appearance of non-monotonicities in close contests among three or more candidates.
Deeper analysis of these effects in three-candidate contests shows that they are closely linked, and in the rare cases where they do occur, tend to occur together.
Their occurrence is also found to reveal either the actual presence of a Condorcet cycle---that unavoidable \emph{sine qua non} of voting paradoxes---or a situation that is `monotonically close' to a Condorcet cycle in that a `silent' shift of preferences in the later part of some voters' ballots converts the election to such a cycle (Proposition \ref{prop:condinstab}).

At the root of AV's more paradoxical outcomes is the fact that ballot rankings appearing below the winner or runner-up in an AV election are `silent': they cannot have any influence on the election outcome for better or worse (Proposition \ref{prop:dompref}).
This basic principle cuts both ways in practice.
On the positive side it ensures that votes for the winner or runner-up cannot possibly be `spoiled' by the effect of later preferences, as can occur with other systems such as Borda count \cite{rg:pppidav}.
On the negative side, it is possible these silent preferences could mask different election outcomes, which thereby become contingent on one candidate rather than another being eliminated at some stage of the count (potentially with only a handful of votes making the difference).
Although close contests are unavoidable in politics, the indirect nature of this effect is a ground for criticism.

The primary reason for wanting to avoid non-monotonicities in a voting system are that they can incentivise tactical voting: strategically casting a ballot contrary to one's true preference in the hope of gaining advantage in some other way, such as through the election of a better supported second-best candidate.
But while the incentive---even necessity---of tactical voting is evident and straightforward in many other systems, including notoriously FPTP, tactical voting in AV even where it is of theoretical benefit is extremely difficult to detect and exploit in practice.
Voters must somehow be able to anticipate the occurrence of a Condorcet cycle or instability, and then contrive to deliver just the right number of votes to certain candidates to achieve a favourable order of elimination.
While the celebrated Gibbard--Satterthwaite theorem states in essence that no voting system can eliminate all opportunities for tactical voting, in the case of AV such opportunities appear rarely effective other than with hindsight.

Finally, one sees that most of the properties of AV also carry across to its STV generalisation to multiple-winner elections.
The benefit that STV offers over party-list systems for such elections stems from its dedication to the principle of voter choice: it is voters themselves rather than political parties that determine the proportional allocation of winning candidates according to political alignment.
At the same time, when voters do rank candidates in a disciplined manner to form stable coalitions, STV resembles in most respects a party-list system on the greatest-remainder principle.

Australian voting systems have evolved in a unique manner, and as has hopefully been shown, their study can provide a valuable guide for the development of democratic voting systems globally.

\appendix

\section{Calculations for quasi-impartial culture model}

Note on nomenclature: in all formulae presented here, `$\log x$' denotes the natural logarithm of $x$.

\subsection{Weighted area $\rareap$ for profiles satisfying Proposition \ref{prop:abprofiles}}
\label{sec:rareap}

Under the quasi-impartial culture model, the second preferences in a three-candidate election are assumed to follow a uniform distribution; accordingly to satisfy condition 2 of Proposition \ref{prop:abprofiles}, the primary vote profile described by the pair $(\rema, \remb)$ with $\rema > 0$ should be weighted by the fraction
\[\weight{\rema}{\remb} = \frac{\remb}{\rema + \remb} \qquad (\rema \geq 0)\]
and the corresponding weighted area required for equation (\ref{eq:abarea}) in the main text is
\begin{multline}
\rareap = \iint_{\rega, \rema > 0} \weight{\rema}{\remb} d\rema d\remb \\
   = \int_0^{1/4} d\remb \int_0^{f(\remb)} \frac{\remb}{\rema + \remb} d\rema \qquad
   f(\remb) = \begin{cases} 1/4 - \remb / 2 & \remb < 1/6 \\ 1/2 - 2 \remb & \remb > 1/6 \end{cases}.
\label{eq:abareap0}
\end{multline}
Here $f(\remb)$ is the value of $\rema$ for which candidate $C$ ties with either $A$ or $B$---equivalently, the upper boundary of the set in Figure \ref{fig:winloseset}---for a given value of the ordinate $\remb$ as that proceeds from its minimum (zero) to its maximum value ($1/4$).
For all values $\remb > 0$ and $0 < \rema < f(\remb)$, the profile $(\rema, \remb)$ is one for which $(A,B)$ is the dominant pair, and the winner and runner-up depend on the preferences of eliminated candidate $C$---hence the weighting $\weight{\rema}{\remb}$.

This integral (\ref{eq:abareap0}) may in turn be evaluated as
\begin{equation}
\begin{split}
\rareap &= \int_0^{1/4} \remb d\remb \int_0^{f(\remb)} \frac{d\rema}{\rema + \remb}
   = \int_0^{1/4} \remb \log\paren{1 + \frac{f(\remb)}{\remb}} d\remb \\
   &= \int_0^{1/6} \remb \log\paren{\half + \frac{1}{4 \remb}} d\remb
      + \int_{1/6}^{1/4} \remb \log\paren{\frac{1}{2\remb} - 1} d\remb \\
   &= \int_2^{\infty} \frac{\log u}{2 \paren{2u - 1}^3} du + \int_1^2 \frac{\log u}{4 \paren{u + 1}^3} du.
\end{split}
\label{eq:abareap1}
\end{equation}
The last of these steps is accomplished with a change of variable, writing $u$ for the argument of the natural logarithm in each case.

Direct evaluation of these integrals is laborious, but the classic reference \cite[\S 2.727 4]{gr:toisp} furnishes the formula
\begin{equation}
\int \frac{\log x}{\paren{b x + a}^3} dx
   = -\frac{\log x}{2b \paren{bx + a}^2} + \frac{1}{2ab \paren{bx + a}}
      + \frac{1}{2 a^2 b} \log\frac{x}{bx + a}.
\label{eq:logx3int}
\end{equation}
Noting that
\begin{equation}
\lim_{x \goesto \infty} \frac{\log x}{x^k} = 0, \qquad
\lim_{x \goesto \infty} \frac{1}{x^k} = 0 \text{ for } k > 0, \qquad \text{and} \qquad
\lim_{x \goesto \infty} \log\frac{x}{bx + a} = \log\frac{1}{b} = - \log b,
\label{eq:loglimits}
\end{equation}
one may thus evaluate
\begin{align}
\rareap &= \half \paren{-\frac{1}{4} \log 2 + \frac{1}{36} \log 2 + \frac{1}{12} - \frac{1}{4} \log\frac{2}{3}}
   + \frac{1}{4} \paren{-\frac{1}{18} \log 2 + \frac{1}{6} + \half \log\frac{2}{3} - \frac{1}{4} + \half \log 2}
   \nonumber \\
   &= \frac{1}{24} + \frac{1}{24} - \frac{1}{16} = \frac{1}{48}.
\label{eq:abareap2}
\end{align}
Thus the weighted area $\rareap$ is $1 / 48$, as asserted in the main text.

\subsection{Weighted area $\twarea$ for profiles satisfying Proposition \ref{prop:winlose}}
\label{sec:twarea}

This appendix deals with the calculation of the weighted area integral
\[\twarea = \iint_{\triwbl} \weight{\rema}{\remb} d\rema d\remb\]
where $\triwbl$ is the triangular set depicted in Figure \ref{fig:winloseset}, and $\weight{\rema}{\remb}$ is given by (\ref{eq:wblweight}).
Notice that if the latter is written as
\begin{equation}
\weight{\rema}{\remb} = \frac{\remb}{1/2 - \remb} \cdot \frac{\rema}{\rema + \remb},
\label{eq:wblweightb}
\end{equation}
the first term depends on $\remb$ alone and may therefore be separated in the integral, which becomes
\begin{equation}
\twarea = \int_0^{1/4} \frac{\remb}{1/2 - \remb} d\remb
   \int_{1/4 - \remb}^{f(\remb)} \frac{\rema}{\rema + \remb} d\rema \qquad
   f(\remb) = \begin{cases} 1/4 - \remb / 2 & \remb < 1/6 \\ 1/2 - 2 \remb & \remb > 1/6 \end{cases}.
\label{eq:twarea0}
\end{equation}
Here $f(\remb)$ is the same function as in (\ref{eq:abareap0}) above, and describes the upper limit of $\rema$ at the shorter `top' edges of the set $\triwbl$, while $1/4 - \remb$ describes the lower limit of $\rema$ at the longer `bottom' edge.
One may then calculate
\begin{equation}
\begin{split}
\twarea &= \int_0^{1/4} \frac{\remb}{1/2 - \remb} d\remb
      \int_{1/4 - \remb}^{f(\remb)} \paren{1 - \frac{\remb}{\rema + \remb}} d\rema \\
   &= \int_0^{1/4} \frac{\remb}{1/2 - \remb} d\remb
      \paren{f(\remb) + \remb - \frac{1}{4} - \remb \log\paren{\frac{f(\remb) + \remb}{1/4}}} \\
   &= \int_0^{1/6} \frac{\remb^2}{1/2 - \remb} \paren{\half - \log\paren{1 + 2 \remb}} d\remb
      \\&\qquad\qquad + \int_{1/6}^{1/4} \frac{\remb}{1/2 - \remb}
         \paren{\frac{1}{4} - \remb - \remb \log\paren{2 - 4 \remb}} d\remb \\
   &= \twarea[1] + \twarea[2].
\end{split}
\label{eq:twarea12}
\end{equation}
The evaluation of these two integrals will follow, after recapitulating some general formulae that will assist in this more extensive exercise.

\subsubsection{General integration formulae}

The following formulae, involving routine calculations only, assist in evaluating $\twarea[1]$ and $\twarea[2]$ above:
\begin{alignat}{2}
\int_a^b \frac{x}{1/2 - x} dx &= \int_a^b \paren{\frac{1}{1 - 2x} - 1} dx
   &&= \half \log\frac{1/2 - a}{1/2 - b} - \paren{b - a},
\label{eq:intxox} \\
\int_a^b \frac{x^2}{1/2 - x} dx &= \int_a^b \paren{\frac{x}{1 - 2x} - x} dx
   &&= \frac{1}{4} \log\frac{1/2 - a}{1/2 - b} - \frac{b - a}{2} \paren{1 + a + b}.
\label{eq:intx2ox}
\end{alignat}
Meanwhile, the integrals involving logarithms are cases of the following formula:
\begin{equation}
\int_a^b \frac{x^2}{1/2 - x} \log\paren{1 \pm 2x} dx
   = - \frac{1}{4} \int_{1 \pm 2a}^{1 \pm 2b} \frac{\paren{u - 1}^2}{u - \paren{1 \pm 1}} \log u \: du.
\label{eq:intlogx2ox}
\end{equation}
With one exception, all terms in the right hand integral in (\ref{eq:intlogx2ox}) can be evaluated using the following standard antiderivatives (omitting additive constants):
\begin{equation}
\int \log u \: du = u \paren{\log u - 1}, \quad
\int u \log u \: du = \frac{u^2}{2} \paren{\log u - \half}, \quad \text{and} \quad
\int \frac{\log u}{u} du = \frac{\paren{\log u}^2}{2}.
\label{eq:logantid}
\end{equation}
The exception to this occurs in the positive case of (\ref{eq:intlogx2ox}), where one encounters the integral of $\log u / (u - 2)$.
There is no expression for this using elementary functions, and one must resort to a special function as below.

\subsubsection{Spence's function or dilogarithm $\plog(z)$}

The analytic function $\plog(z)$, known as either Spence's function or the dilogarithm \cite[\S 6.15]{z:csmtf}, is a special function defined by either of the equivalent formulae
\begin{equation}
\plog(z) = - \int_0^z \frac{\log\paren{1 - t}}{t} dt = \int_1^{1 - z} \frac{\log t}{1 - t} dt.
\label{eq:spence}
\end{equation}
(Using the second form of the definition, some sources instead write $\mathop{\mathrm{dilog}}(z) = \plog(1 - z)$.)
For values $|z| \leq 1$, which are the only ones required for present purposes, $\plog(z)$ can be evaluated using the power series
\begin{equation}
\plog(z) = \sum_{k=1}^{\infty} \frac{z^k}{k^2} \qquad \paren{|z| \leq 1}.
\label{eq:spencepow}
\end{equation}
Thus in particular $\plog(0) = 0$, $\plog(z)$ is real for $z \in [-1,1]$ and positive for $z \in (0,1]$, and $\plog(1) = \sum_k (1 / k^2) = \pi^2 / 6$.
Another special value with a closed-form representation is
\begin{equation}
\plog\paren{\half} = \frac{\pi^2}{12} - \frac{\paren{\log 2}^2}{2} \approx 0.582240526465.
\label{eq:spencehalf}
\end{equation}
($\plog(z)$ is one of a general class of \emph{polylogarithms} $\plog[p](z)$ defined recursively by $\plog[1](z) = -\log\paren{1 - z}$ and $\plog[p+1](z) = \int_0^z \plog[p](z) / z$, and having the series representation $\plog[p](z) = \sum_k z^k / k^p$ for $|z| \leq 1$.)

With the aid of the function $\plog(z)$, one may evaluate the integral
\begin{equation}
\int_a^b \frac{\log u}{u - 2} du = - \int_{a/2}^{b/2} \frac{\log 2 + \log t}{1 - t} dt
   = \log 2 \cdot \log\frac{1 - b/2}{1 - a/2} + \plog\paren{1 - \frac{a}{2}} - \plog\paren{1 - \frac{b}{2}}
\label{eq:intlogum}
\end{equation}
which figures in the evaluation of (\ref{eq:intlogx2ox}) in the positive case.
Indeed, the full formulae for both cases of (\ref{eq:intlogx2ox}) may now be determined, with the aid of (\ref{eq:logantid}) and (\ref{eq:intlogum}), as
\begin{multline}
\int_a^b \frac{x^2}{1/2 - x} \log\paren{1 + 2x} dx
   = - \frac{1}{4} \int_{1+2a}^{1+2b} \paren{u + \frac{1}{u - 2}} \log u \: du \\
   = \frac{b-a}{4} \paren{1+a+b} + \frac{a(1+a)}{2} \log\paren{1+2a} - \frac{b(1+b)}{2} \log\paren{1+2b} \\
      + \frac{1}{8} \log\frac{1/2 + a}{1/2 + b} + \frac{\log 2}{4} \log\frac{1/2 - a}{1/2 - b}
      + \frac{1}{4} \brak{\plog\paren{\half - b} - \plog\paren{\half - a}}
\label{eq:intlogpx2ox}
\end{multline}
and
\begin{multline}
\int_a^b \frac{x^2}{1/2 - x} \log\paren{1 - 2x} dx
   = - \frac{1}{4} \int_{1-2a}^{1-2b} \paren{u - 2 + \frac{1}{u}} \log u \: du \\
   = \frac{b-a}{4} \paren{3+a+b} + \frac{a(1+a)}{2} \log\paren{1-2a} - \frac{b(1+b)}{2} \log\paren{1-2b} \\
      + \frac{1}{8} \log\frac{1/2 - b}{1/2 - a} \bparen{3 - \log\brak{\paren{1 - 2a} \paren{1 - 2b}}}.
\label{eq:intlognx2ox}
\end{multline}

\subsubsection{Calculation of $\twarea[1]$, $\twarea[2]$ and $\twarea$}

With the aid of the integration formulae (\ref{eq:intxox}), (\ref{eq:intx2ox}), (\ref{eq:intlogpx2ox}) and (\ref{eq:intlognx2ox}), one may now complete the calculation of $\twarea[1]$ and $\twarea[2]$ as given by (\ref{eq:twarea12}).
For $\twarea[1]$, one has
\begin{equation}
\twarea[1] = \half \int_0^{1/6} \frac{\remb^2}{1/2 - \remb} d\remb
   - \int_0^{1/6} \frac{\remb^2}{1/2 - \remb} \log\paren{1 + 2 \remb} d\remb
\label{eq:twarea1a}
\end{equation}
which may be recognised as cases $a = 0$, $b = 1/6$ of (\ref{eq:intx2ox}) and (\ref{eq:intlogpx2ox}), giving after substitution
\begin{equation}
\twarea[1] = \paren{\frac{1}{8} - \frac{\log 2}{4}} \log\frac{3}{2}
   + \paren{\frac{7}{72} + \frac{1}{8}} \log\frac{4}{3} - \frac{7}{72}
   + \frac{1}{4} \brak{\plog\paren{\half} - \plog\paren{\frac{1}{3}}}.
\label{eq:twarea1b}
\end{equation}
The special value $\plog(1/2)$ is given by (\ref{eq:spencehalf}); unfortunately $\plog(1/3)$ has no known closed form, but evaluation of the power series (\ref{eq:spencepow}) shows that
\begin{equation}
\plog\paren{\frac{1}{3}} \approx 0.366213229977.
\label{eq:spencethird}
\end{equation}
After separating out the logarithms in (\ref{eq:twarea1b}) one obtains the explicit formula for $\twarea[1]$
\begin{equation}
\begin{split}
\twarea[1] &= \frac{23}{72} \log 2 - \frac{7}{72} \log 3 - \frac{\log 2 \log 3}{4} + \frac{\paren{\log 2}^2}{8}
      - \frac{7}{72} + \frac{\pi^2}{48} - \frac{1}{4} \plog\paren{\frac{1}{3}} \\
   &\approx 0.001135340722.
\end{split}
\label{eq:twarea1}
\end{equation}
Now for $\twarea[2]$, by separating $\log\paren{2 - 4 \remb} = \log 2 + \log\paren{1 - 2 \remb}$ one has
\begin{multline}
\twarea[2] =  \frac{1}{4} \int_{1/6}^{1/4} \frac{\remb}{1/2 - \remb} d\remb
   - \paren{1 + \log 2} \int_{1/6}^{1/4} \frac{\remb^2}{1/2 - \remb} d\remb \\
   - \int_{1/6}^{1/4} \frac{\remb^2}{1/2 - \remb} \log\paren{1 - 2 \remb} d\remb.
\label{eq:twarea2a}
\end{multline}
These three terms may be recognised as cases $a = 1/6$ and $b = 1/4$ of (\ref{eq:intxox}), (\ref{eq:intx2ox}) and (\ref{eq:intlognx2ox}), respectively.
After substitution, one obtains
\begin{equation}
\twarea[2] = \frac{1}{8} \log\frac{4}{3}
   - \paren{1 + \log 2} \paren{\frac{1}{4} \log\frac{4}{3} - \frac{17}{288}}
   - \frac{53}{576} - \frac{7}{72} \log\frac{2}{3} - \frac{5}{32} \log 2
   - \frac{1}{8} \log\frac{3}{4} \paren{3 + \log 3}.
\label{eq:twarea2b}
\end{equation}
After expanding and separating out the logarithms, there results
\begin{equation}
\begin{split}
\twarea[2] &= \frac{11}{36} \log 2 - \frac{11}{72} \log 3 + \frac{\log 2 \log 3}{2}
      - \frac{\paren{\log 2}^2}{2} - \frac{\paren{\log 3}^2}{8} - \frac{19}{576} \\
   &\approx 0.000620194773.
\end{split}
\label{eq:twarea2}
\end{equation}
The weighted area $\twarea$ sought according to (\ref{eq:twarea0}) and (\ref{eq:twarea12}) is given by $\twarea[1] + \twarea[2]$, and by combining (\ref{eq:twarea1}) and (\ref{eq:twarea2}) is seen to have the expression given by (\ref{eq:twarea}) in the main text.

\subsection{Weighted area $\tlarea$ for profiles satisfying Proposition \ref{prop:losewin}}
\label{sec:tlarea}

In the same vein as Appendix \ref{sec:twarea} above, this appendix calculates the weighted area
\[\tlarea = \iint_{\trilbw} \weight{\rema}{\remb} d\rema d\remb\]
where $\trilbw$ is the right triangular set depicted in Figure \ref{fig:losewinset}, and $\weight{\rema}{\remb}$ is given by (\ref{eq:lbwweight}), expressed here as
\[\weight{\rema}{\remb} = \frac{\rema}{1/2 - \rema} \cdot \frac{\remb}{\rema + \remb}.\]
Presented this way, the first term depends on $\rema$ alone and, as before, may be separated in the integral, which becomes
\begin{equation}
\tlarea = \int_{1/6}^{1/4} \frac{\rema}{1/2 - \rema} d\rema
   \int_0^{1/2 - 2\rema} \frac{\remb}{\rema + \remb} d\remb.
\label{eq:tlarea0}
\end{equation}
Now proceeding analogously to the calculation of $\twarea$ in Appendix \ref{sec:twarea}, one calculates
\begin{equation}
\begin{split}
\tlarea &= \int_{1/6}^{1/4} \frac{\rema}{1/2 - \rema} d\rema
      \int_0^{1/2 - 2\rema} \paren{1 - \frac{\rema}{\remb + \rema}} d\remb \\
   &= \int_{1/6}^{1/4} \frac{\rema}{1/2 - \rema} d\rema
      \paren{\half - 2 \rema - \rema \log\paren{\frac{1/2 - \rema}{\rema}}} \\
   &= \half \int_{1/6}^{1/4} \frac{\rema}{1/2 - \rema} d\rema
      - 2 \int_{1/6}^{1/4} \frac{\rema^2}{1/2 - \rema} d\rema
      - \int_{1/6}^{1/4} \frac{\rema^2}{1/2 - \rema} \log\paren{\frac{1}{2 \rema} - 1} d\rema.
\end{split}
\label{eq:tlarea1}
\end{equation}
The first two terms here are instances of the general formulae (\ref{eq:intxox}) and (\ref{eq:intx2ox}) in Appendix \ref{sec:twarea}, while the third (now labelled $-\tlarea[X]$ for convenience) resembles an integral calculated at (\ref{eq:abareap1}) in Appendix \ref{sec:rareap}.
As was done there, setting $u$ equal to the argument of the logarithm brings it into the form
\begin{equation}
\tlarea[X] = \int_{1/6}^{1/4} \frac{\rema^2}{1/2 - \rema} \log\paren{\frac{1}{2 \rema} - 1} d\rema
   = \int_1^2 \frac{\log u}{4 u \paren{u + 1}^3} du.
\label{eq:tlareau}
\end{equation}
This is almost an instance of the formula (\ref{eq:logx3int}) from Appendix \ref{sec:rareap}, but for the additional factor $u$ in the denominator.
To deal with this, one may resort to partial fractions, and the identity
\begin{equation}
\frac{1}{u \paren{u + 1}^3}
   = \frac{1}{u} - \frac{1}{u + 1} - \frac{1}{\paren{u + 1}^2} - \frac{1}{\paren{u + 1}^3},
\label{eq:uu3pf}
\end{equation}
which allows $\tlarea[X]$ to be evaluated as follows:
\begin{equation}
4 \tlarea[X] = \int_1^2 \frac{\log u}{u} du - \int_1^2 \frac{\log u}{u + 1} du
   - \int_1^2 \frac{\log u}{\paren{u + 1}^2} du - \int_1^2 \frac{\log u}{\paren{u + 1}^3} du.
\label{eq:tlax}
\end{equation}
Of these four integrals, the first is straightforward using (\ref{eq:logantid}) in Appendix \ref{sec:twarea}; the fourth is obtained using (\ref{eq:logx3int}) as already noted; while the third is obtained from the formula \cite[\S 2.727 3]{gr:toisp}
\begin{equation}
\int \frac{\log x}{\paren{bx + a}^2} dx = - \frac{\log x}{b \paren{bx + a}} + \frac{1}{ab} \log\frac{x}{bx + a}.
\label{eq:logx2int}
\end{equation}
This leaves the second integral in (\ref{eq:tlax}), which is analysed as follows:
\begin{multline}
\int_1^2 \frac{\log u}{u + 1} du = \int_2^3 \frac{\log\paren{t - 1}}{t} dt
   = \int_2^3 \frac{\log t + \log\paren{1 - 1/t}}{t} dt
   = \int_2^3 \frac{\log t}{t} dt - \int_{1/2}^{1/3} \frac{\log\paren{1 - x}}{x} dx \\
   = \frac{\paren{\log 3}^2}{2} - \frac{\paren{\log 2}^2}{2} + \plog\paren{\frac{1}{3}} - \plog\paren{\half}
   = \frac{\paren{\log 3}^2}{2} - \frac{\pi^2}{12} + \plog\paren{\frac{1}{3}},
\label{eq:logx1int}
\end{multline}
where $\plog(z)$ is the Spence function introduced in Appendix \ref{sec:twarea}.
Putting this all together,
\begin{equation}
\begin{split}
4 \tlarea[X]
   &= \frac{\paren{\log 2}^2}{2} - \frac{\paren{\log 3}^2}{2} + \frac{\pi^2}{12} - \plog\paren{\frac{1}{3}} \\
   &\qquad - \paren{- \frac{\log 2}{3} + \log\frac{2}{3} - \log\frac{1}{2}} - \paren{- \frac{\log 2}{18}
      + \frac{1}{6} + \frac{1}{2} \log\frac{2}{3} - \frac{1}{4} - \frac{1}{2} \log\frac{1}{2}} \\
   &= \frac{1}{12} - \frac{47}{18} \log 2 + \frac{3}{2} \log 3 + \frac{\paren{\log 2}^2}{2}
      - \frac{\paren{\log 3}^2}{2} + \frac{\pi^2}{12} - \plog\paren{\frac{1}{3}}.
\end{split}
\label{eq:tlax2}
\end{equation}
Finally, one calculates the original area $\tlarea$ required as
\begin{equation}
\begin{split}
\tlarea &= \frac{11}{144} - \frac{1}{4} \log\frac{4}{3} - \tlarea[X] \\
   &= \frac{1}{18} + \frac{11}{72} \log 2 - \frac{\log 3}{8} - \frac{\paren{\log 2}^2}{8}
      + \frac{\paren{\log 3}^2}{8} - \frac{\pi^2}{48} + \frac{1}{4} \plog\paren{\frac{1}{3}}
\end{split}
\label{eq:tlarea2}
\end{equation}
as stated in the main text.

\section{Proof of Proposition \ref{prop:condinstab}}
\label{sec:condinstab}

The purpose of this appendix is to show that in a winner-takes-all election, if there is a Condorcet winner $C$ who is not elected by AV, then the voting profile can be modified, without altering any primary votes, to convert the election to a Condorcet cycle.
Further, this can be done in a way that does not alter the election outcome or the result at any stage of the AV counting process.

First recall that if $C$ is the Condorcet winner but is not elected then there is a dominant pair comprising two candidates $(A, B)$ distinct from $C$ of which one (say $A$) is elected.
$C$ meanwhile must by Proposition \ref{prop:condorcet} have one-third or fewer votes at every stage of counting.
One then has $A \pref B$ by the dominant-pair property, while $C \pref A$ and $C \pref B$ since $C$ is the Condorcet winner.

These dominance relations will instead form a Condorcet cycle if the single dominance relation between $B$ and $C$ can be reversed from $C \pref B$ to $B \pref C$ without affecting the other two relations.
This will necessarily also be a `top cycle' (one not itself dominated), since no other candidate is preferred by a majority to $C$.

The first (and perhaps only) step required to accomplish this reversal is to consider those ballots that include the preference ordering $A > C > B$, and exchange the ranking of $B$ and $C$ within these ballots.
If one temporarily imagines all candidates other than $(A, B, C)$ to be eliminated, this amounts to shifting ballots from profile type $\alpha$ to type $a$ in Table \ref{tab:types3}.
Note that by virtue of Proposition \ref{prop:dompref} this will have no effect on the count or on the election outcome since $A$ is one of the dominant pair, and will also have no effect on the dominance relations involving $A$.

In scenarios where Condorcet winner $C$ is the \emph{last} candidate to be eliminated prior to the dominant pair---that is, the AV procedure ultimately reduces to a three-candidate contest between $A$, $B$ and $C$---this exchange of $B$ and $C$ preferences in `type $\alpha$' ballots suffices by itself to establish the Condorcet cycle.
This is because $C$ can have at most one-third of votes at this stage, and all other ballots can be made to prefer $B$ to $C$.
This establishes Proposition \ref{prop:condinstab} in those cases where $C$ is eliminated last before the dominant pair, including in all three-candidate elections.

A complication arises however when $C$ is eliminated earlier in the process, and the eventual three-way contest is between $(A, B, D)$ for some candidate $D$ other than $C$.
It may then eventuate that $A$ and $B$ themselves have attained their dominant-pair position through preference flows from $C$, so that ballots can at this stage appear as votes for either $A$ or $B$, despite actually preferring $C$ and therefore contributing to the relation $C \pref B$.
When these are combined with ballots having the preference order $D > C > B$, there can be a majority of ballots preferring $C$ to $B$ irrespective of how the remaining $A$ ballots are altered.

Table \ref{tab:condor4} provides an example with four candidates and 90 voters illustrating this difficulty.
\begin{table}
\begin{centre}
\begin{tabular}{r|r}
Ballot order & Votes \\ \hline
$A > C > B > D$ & 22 \\
$B > C > A > D$ & 21 \\
$C > A > B > D$ & 10 \\
$C > B > A > D$ & 9 \\
$D > C > A > B$ & 28 \\ \hline
Total ballots & 90
\end{tabular}
\end{centre}
\caption{Four-candidate AV election with early eliminated Condorcet winner.}
\label{tab:condor4}
\end{table}
It may readily be verified that $C$ is the Condorcet winner, but with only 19 primary votes is the first candidate eliminated.
The resulting three-way contest is between $A$, $B$ and $D$, and although $D$ has fewer than one-third of votes, there are 19 ballots with $A$ and $B$ for which $C$ is preferred and this preference cannot be reversed without altering the primary vote for $C$.
When combined with the $D$ votes, all of which prefer $C$ to $B$, one finds that the relation $C \pref B$ persists and does so even if $B$ and $C$ preferences are reversed in all 22 of $A$'s votes.

However, the resolution of this difficulty is also evident from this example, as the option remains open to modify the $D$ ballots by shifting $C$ to below $B$ in the preference ordering, without altering the relative position of any other candidates.
Once again, such a shift does not alter any primary votes, and because $C$ has already been eliminated, does not affect the election outcome or the counting process.
This may be done in combination with the shift in $A$ preferences above to ensure $B$ is preferred to $C$ overall.

Were the entire $D$ vote shifted to prefer $B$ to $C$---by changing all $D > C > A > B$ ballots to $D > A > B > C$---this would more than suffice to ensure $B \pref C$, since the only ballots now preferring $C$ to $B$ are those that place their primary vote with $C$, and since $C$ is eliminated ahead of the three-way contest between $(A, B, D)$, this is necessarily less than one-quarter of all ballots.
However, the example also shows that to change \emph{all} $D$ ballots in this way also risks reversing the relation $C \pref A$ and turning $A$ into the Condorcet winner.
While this would conveniently align the outcome of the AV election with that approved by the good Marquis, it is not the outcome promised under Proposition \ref{prop:condinstab}.

In the example of Table \ref{tab:condor4}, it suffices to move $C$ to the bottom of at least 3 or at most 22 out of 28 ballots cast for $D$, and together with exchanging $B$ and $C$ in the 22 ballots for $A$, this ensures that $B \pref C$ while preserving the relations $C \pref A$ and $A \pref B$ to complete the Condorcet cycle.
It remains to show that this can always be done in general.

Consider the stage of the AV process where $C$ is eliminated, and suppose there are four or more contesting candidates at this stage.
(The case of three candidates was covered above.)
Let $\chi \geq 0$ be the proportion of votes held by $C$ at this stage: thus, transferring these votes from $C$ forces $C > B$ in this share $\chi$ of all ballots.
Meanwhile, the shares of votes that lie with $A$ or $B$ before this transfer, and can therefore be modified if necessary to have $B > C$ in the ballot order, are denoted $\alpha$ and $\beta$ respectively; each of these must clearly be strictly greater than $\chi$.

The `remainder' vote share, a fraction $1 - \chi - \alpha - \beta$, sits with other candidates and has these candidates placed ahead of $A$, $B$ and $C$ in their preference ordering.
Assuming $\alpha + \chi < 1/2$ for now, a share $1/ 2 - \alpha - \chi$ of these `remainder' ballots, at least, must prefer $C$ to $B$ in order to have $C \pref B$ overall.
All such ballots may be modified to place $C$ below $B$ in their preferences without altering the outcome of the AV process (since $C$ is eliminated before this preference counts).
However, for any ballots that contain the preference ordering $C > A > B$ (as in the Table \ref{tab:condor4} example), this has the unwanted side effect of reversing the preference for $C$ over $A$.

The most severe case is where \emph{all} such ballots have the $C > A > B$ ordering.
In this case, any shift of a fraction $\delta \leq 1/2 - \alpha - \chi$ of ballots to place $C$ below $B$ will also reduce the share of ballots preferring $C$ to $A$ by the same fraction $\delta$.

After shifting these ballots, and modifying the ballots with preference order $A > C > B$ to have order $A > B > C$ instead, the proportion still having $C$ above $B$ in the preference ordering can be reduced to just those transferred from $C$, and those of the `remainder' ballots not altered, or
\begin{equation}
(C > B)_{\min} = \chi + (1 - \chi - \alpha - \beta - \delta) = 1 - \alpha - \beta - \delta
\label{eq:cbmin}
\end{equation}
and is less than $1/2$ provided
\begin{equation}
\delta > \half - \alpha - \beta.
\label{eq:cbdmin}
\end{equation}
Meanwhile, the proportion of ballots having $A$ above $C$ in the preference ordering can be minimised by swapping $C$ and $A$ in all ballots that contain the preference order $B > A > C$ (note these may be ballots cast for $B$, or `remainder' ballots with some other candidate in first place). Again, since $B$ is part of the dominant pair this does not change the outcome of the AV count, as a consequence of Proposition \ref{prop:dompref}.
In this way the proportion of ballots with $A$ above $C$ can be brought as low as $\alpha + \delta$, where again $\alpha > \chi$ is the share of ballots cast for $A$ before $C$ is eliminated.
To keep this to a minority, it therefore suffices that
\begin{equation}
\delta < \half - \alpha.
\label{eq:cbdmax}
\end{equation}
It follows that when $\alpha + \chi < 1/2$, there is a nonempty interval of values $\delta \in (1/2 - \alpha - \beta, 1/2 - \alpha - \chi)$, such that out of the `remainder' ballots at the point where $C$ is eliminated that have $C$ preferred to $B$ (having a share at least $1/2 - \alpha - \chi \leq 1/2 - \alpha$ of all votes), a share $\delta$ of ballots can be modified to place $C$ below $B$, \emph{and} also below $A$ if necessary, to ensure $B \pref C$ while preserving the relations $C \pref A$ and $A \pref B$.

Lastly, it is necessary to consider the case $\alpha + \chi \geq 1/2$, where it is possible to have \emph{no} `remainder' bellots preferring $C$ to $B$, yet still have $C \pref B$ before any modifications.
But in this case, the bound (\ref{eq:cbmin}) even with $\delta = 0$ evaluates to
\begin{equation}
(C > B)_{\min} = 1 - \alpha - \beta \leq \half - \paren{\beta - \chi} < \half,
\label{eq:cbmin0}
\end{equation}
and so no alterations to the `remainder' ballots are necessary in this case to reverse the $C \pref B$ relation.

This establishes Proposition \ref{prop:condinstab} in the general case.
To recap, under the assumption that Condorcet winner $C$ is not elected, so that the winner $A$ and runner-up $B$ are distinct from $C$, the following ballot changes suffice to induce a top Condorcet cycle:
\begin{enumerate}
\item
In ballots that contain the preference ordering $A > C > B$, swap $B$ and $C$ in the ballot order.
\item
In ballots that contain the preference ordering $B > A > C$, swap $A$ and $C$ in the ballot order.
\item
If when $C$ is eliminated there remain contesting candidates other than $A$ and $B$, let $N$ be the total number of ballots, $a > 0$ the number counting for $A$ and $b > 0$ the number counting for $B$ at this stage (before distributing preferences from $C$).
Set $d = \floor{N / 2} + 1 - a - b$.
Then if $d > 0$, there will be (in the original scenario) at least $d$ ballots that count for candidates other than $A$, $B$ and $C$ at this stage \emph{and} that rank $C$ above $B$ in their subsequent preferences.
In any $d$ of such ballots, move $C$ to rank below $B$ (if this was not already done at step 1) without altering the relative position of any other candidates.
If $d \leq 0$, make no further alterations.
\end{enumerate}
If Condorcet winner $C$ is eliminated only at the point where $A$ and $B$ are the only other contesting candidates, only the first of these changes is necessary.
None of these changes will have any effect on primary votes, on the vote tally at any stage of the AV counting process, or on the ultimate election outcome, as they only affect dominant-pair preferences that are silent under Proposition \ref{prop:dompref}, or the relative position of $C$ following $C$'s elimination.

\bibliographystyle{plain}
\bibliography{voting}

\end{document}